\documentclass[fleqn,usenatbib]{mnras}

\usepackage{newtxtext,newtxmath}

\usepackage[T1]{fontenc}
\usepackage{ae,aecompl}
\usepackage{ulem}

\usepackage{graphicx}	
\usepackage{amsmath}	

\newcommand{\pluto}{\textsc{pluto }}
\newcommand{\cloudy}{\textsc{cloudy }}
\newcommand{\strom}{Str\"{o}mgren } 

\newcommand{\ps}{\mbox{ s}^{-1} }
\newcommand{\cc}{\mbox{ cm}^{3} }
\newcommand{\pcc}{\mbox{ cm}^{-3} }
\newcommand{\pcm}{\mbox{ cm}^{-2} }	

\newcommand{\ergps}{\mbox{ erg s}^{-1} }
\newcommand{\ergpspcmsq}{\mbox{ erg s}^{-1} \mbox{cm}^{-2} }
\newcommand{\deltaray}{\textit{delta ray }}
\newcommand{\gaussbeam}{\textit{Gaussian beam }}

\title[ionisation and radiation in \pluto]{A new ionisation network and radiation transport module in \pluto}

\author[Sarkar, Sternberg \& Gnat]{
Kartick C. Sarkar,$^{1}$\thanks{E-mail: sarkar.kartick@mail.huji.ac.il, kartick.c.sarkar100@gmail.com}
Amiel Sternberg$^{2,3,4}$
and Orly Gnat $^{1}$
\\
$^{1}$Racah Institute of Physics, The Hebrew University of Jerusalem, 91904, Israel\\
$^{2}$School of Physics and Astronomy, Tel Aviv University, Ramat Aviv, 69978, Israel\\
$^{3}$Centre for Computational Astrophysics, Flatiron Institute, 162 5th Avenue, 10010, New York, NY, USA\\
$^{4}$Max-Planck-Institut fur Extraterrestrische Physik (MPE), Giessenbachstr., 85748, Garching, FRG
}

\date{Accepted XXX. Received YYY; in original form ZZZ}

\pubyear{2020}

\begin{document}
\label{firstpage}
\pagerange{\pageref{firstpage}--\pageref{lastpage}}
\maketitle

\begin{abstract}
We introduce a new general-purpose time-dependent ionisation network (IN) and a radiation transport (RT) module for the magneto-hydrodynamic (MHD) code \pluto. Our ionisation network is reliable for temperatures ranging from $5\times 10^3$ to $3\times 10^8$~K and includes all ionisation states of H, He, C, N, O, Ne, Mg, Si, S and Fe, making it suitable for studying a variety of astrophysical scenarios. Radiation loss for each ion-electron pair is calculated using \cloudy-17 data on-the-fly. Photo-ionisation and charge exchange are the main processes contributing to chemical heating. The IN is fully coupled to the radiation transport module over a large range of opacities at different frequencies. The RT module employs a method of short characteristics assuming spherical symmetry. 
The radiation module requires the assumption of spherical symmetry, while the IN is compatible with full 3D. We also include a simple prescription for dust opacity, grain destruction, and the dust contribution to radiation pressure. We present numerical tests to show the reliability and limitations of the new modules. We also present a post-processing tool to calculate projected column densities and emission spectra.  
\end{abstract}


\section{Introduction}
Electromagnetic line and continuum radiation from ionised plasma are critical diagnostic probes of the underlying physical mechanisms operating in astrophysical environments.
Modelling radiation transport is, therefore, an important ingredient in astrophysical plasma simulations.
 Most hydrodynamic plasma models 
focus on the kinetic and thermal conditions,
with less attention on the actual ionisation states of the heavy element constituents that enable dynamically important energy losses \citep{Cunningham2005, Stone2008, Jiang2012, Rosdahl2013}. The ionisation states are often assumed to be distributed in equilibrium configurations depending only on the local temperature and/or density of the gas. Examples of such assumptions include temperature-dependent collisional ionisation equilibrium (CIE) or density-dependent photo-ionisation equilibrium (PIE) in an externally set radiation field. 
However, equilibrium is valid only when the plasma has enough time to fully respond to changes in the thermal energies
or radiation fields.
For CIE, this requires that the ionisation and recombination time-scales, $\tau_{\rm ion}$ and $\tau_{\rm rec}$, are much smaller than the time-scale to change the internal energy $\tau_{\rm th}$. 
If not, the plasma may be under or over-ionised, depending on the time-scale-ratio and the thermal evolution.

For example, for radiatively cooling and recombining gas, 
the radiative cooling time is
\begin{equation}
\tau_{\rm cool} \sim \frac{1.5\: n_0\:k_B\:T}{n_0^2 \Lambda(T, Z)},
\end{equation}
while the recombination time for a given ion, $i$, is 
\begin{equation}
\tau_{\rm rec,i} \sim \frac{1}{n_0\: \alpha_i (T)}
\end{equation}
where, $\alpha_i (T)$ is the total (radiative + dielectronic) recombination rate coefficient. 
The ratio between the cooling and recombination times is then,
\begin{equation}
    \frac{\tau_{\rm cool}}{\tau_{\rm rec,i}}\sim \frac{1.5\:k_B\:T\: \alpha_i (T)}{\Lambda(T, Z)},
\end{equation}
which is independent of density. For example, for an initial temperature of $\sim 10^5$ K, and for the 
\ion{C}{vi}$\rightarrow$\ion{C}{v} recombination, the timescale-ratio is $\sim 0.18$ ($\alpha_{CVI} = 3.5\times 10^{-12} \ps$ and $\Lambda(10^5) = 4 \times 10^{-22} \ergps \cc$).
Therefore, \ion{C}{v} will be over-abundant compared to CIE.  
Clearly, there is a need to consider the time evolution of a non-equilibrium ionisation (NEI) network in addition to the hydrodynamic variables in such cases. 
Many authors have studied the isochoric/isobaric cooling of hot gas from $t \sim 10^6$~K to $\approx 10^4$~K and shown that the time-dependent ion fractions for ions can differ by 
orders of magnitude (which in turn affect gas cooling) indicating the importance of the non-equilibrium conditions \citep{Kafatos1973, Shapiro1976, Schmutzler1993, Gnat2007, Oppenheimer2013, Gnat2017}

Observationally, the presence of non-equilibrium plasma has been shown in many environments. 
Studies including \cite{Becker1980, Claas1989, Brinkmann1999, Bamba2016, Suzuki2018} found evidence for under-ionised/over-ionised plasma in the X-ray spectrum of young ($\lesssim 2000$ yr) SN remnants. 
Corresponding simulations of the supernovae bubbles also show significant effects of the NEI on observable ions \citep{Hamilton1983, Shull1983, Itoh1988, Slavin1992, Slavin2015, ZhangGao2019} . Non-equilibrium effects have also been attributed to the non-detection of  \ion{N}{v} compared to \ion{O}{vi} or failure of a CIE fit in the galactic winds \citep{Breitschwerdt1999, Breitschwerdt2003, Chisholm2018, Gray2019}, excess X-ray background \citep{Breitschwerdt1994}, lack of \ion{C}{iv} and \ion{N}{v} in the local bubble \citep{DeAvillez2012} 
and the missing baryons in the warm-hot ionised medium (WHIM) in the intergalactic medium (IGM) \citep{Yoshikawa2003, Cen2006, Bertone2008}. 

Numerically there have been several
considerations of non-equilibrium ionisation effects. Some of the codes \citep{Kafatos1973, Shapiro1976,Schmutzler1993,Gnat2007,Bradshaw2009,Gnat2017} only studied the temporal evolution of the IN of a plasma that did not involve spatial dynamics. One approximate method to combine spatial dynamics with the ionisation network is to calculate the cooling and heating rates based on isochoric or isobaric evolution and then use these tables in a full hydro calculation \citep{Sutherland2003, Vasiliev2013a}.  More comprehensive 1D steady state codes have been developed to include a self-consistent ionisation network that evolves with fluid dynamics. Among them, \cite{Shull1979, Allen2008, Gnat2009} consider a coupling between the radiative transfer and the ionisation. However, these models are limited to solving a plane parallel steady state shock for a given shock velocity. \cite{Breitschwerdt1999, Slavin1992, Slavin2015} consider solutions in spherical geometry but do not include any radiative transfer.

A popular method of solving the RT problem in a scattering dominated system is to use the Eddington approximation. This method is valid when the specific intensity is assumed to be nearly isotropic, or up to a linear dependence on $\cos(\theta)$ from the direction of propagation \citep{Chandrasekhar1960, Hummer1971,Hummer1973,ryb+light}. However, this method is inadequate in the optically thin limit as the specific intensity becomes strongly forward peaked. Techniques to overcome this problem include using a direct flux limiter in the free streaming limit \citep{Levermore1981} or using the M1 closure method  \citep{Levermore1984, Gnedin2001, Fuksman2019} in which the second and zeroth moment of specific intensity are connected through an Eddington tensor that works both in optically thick 
and thin limits. The specific form of the Eddington tensor is, however, chosen in an \textit{ad-hoc} way. The state-of-the-art method is to use the method of rays to solve for the Eddington tensor at each location and use this tensor for closing the moment equations \citep{Stone1992, Davis2012, Jiang2012}. 

Full 3D MHD codes typically include either IN or RT but rarely both together. The earliest such attempts was made in \textsc{yguazu} \citep{Raga1999, Raga2000} which included a small IN and a radiative transfer. Their technique, unfortunately, is suitable for only a limited number of emitting sources and a constant grid spacing. \textsc{astrobear} \citep{Cunningham2005, Cunningham2009} contains a network of H and He ions but does not consider any RT or metals. \textsc{athena++} \citep{Stone2008,Davis2012, Jiang2012} contains a state-of-the-art radiative transfer module (privately distributed) but does not consider the IN dynamics. While \textsc{flash-fervent} \citep{Fryxell2000, Baczynski2015}, \textsc{ramses-RT} \citep{Rosdahl2013} and \textsc{arepo-rt} \citep{Kannan2019} contain some form of radiative transfer coupled to chemical network, the network contains only few ions/molecules (mostly, H, He, CO etc.) and the metals are assumed to be in photo-ionisation equilibrium. Most of the 3D radiative transfer modules like \textsc{skirt}  \citep{Baes2003}, \textsc{sunrise} \citep{Jonsson2006}, \textsc{hyperion} \citep{Robitaille2011} and \textsc{radmc-3d} \citep{Dullemond2012} that have the capability of including a full ionisation network and dust 
using the Monte-Carlo method, can only be used as post-processing tools due to their massively complex physics and, therefore, slower computation speed. Although \textsc{torus-3dpdr} \citep{Harries2000, Bisbas2015a} solves on-the-fly radiative transfer using the Monte-Carlo technique, it assumes an equilibrium chemistry network.

With our aim of combining both the IN and the RT in a single MHD code, we extend the already existing IN of \pluto  \citep{Mignone2007, Tesileanu2008} \footnote{The \pluto-4.0 includes all ionisation levels of H, He but only till $4+$ ions for C, N, O, Ne and S, and till $3+$ for Fe.} to include all the ionisation states of H, He, C, N, O, Ne, Mg, Si, S and Fe. The network of ionisation states can be reduced if required. Our radiative transfer module uses a discrete ordinate technique (short characteristics) in spherical symmetry where the RT is solved along different rays fixed in space and angles to transport a spectrum.
 Spherical symmetry enables methods that speed up the calculation. Discrete ordinates have been used in some form or other for various purposes ranging from neutrino transport inside a supernova to the neutron transport problem inside nuclear reactors \citep[see for example][]{Hill1975,Lewis1984, Birnboim2000}.  This method does not suffer from the challenges of traversing from an optically thin to an optically thick medium or vice versa. With the inclusion of RT, we also include photo-heating, charge exchange heating/cooling in the IN and radiation pressure on the fluid dynamically calculated at each time step.
 \footnote{The modified version of the code is available in \url{https://gitlab.com/kartickchsarkar/pluto-neq-radiation}.}

In the following sections, we describe the numerical implementation, equations and standard tests to establish that our module is suitable for studying different astrophysical systems.  In a companion paper \citep[][ paper-II]{Sarkar2020b} we make use of this tool to study the time evolution of heavy element column densities in (non steady state) expanding supernova remnant.

\section{The equations}
\subsection{The MHD equations}
\label{subsec:mhd-eqs}
The MHD equations and numerical implementation of the ionisation network are as described in \cite{Tesileanu2008}. We added some more ions and extended this network in terms of new reactions which will be shortly discussed. The ionisation module is suitable for the temperature range of $5\times 10^3 \leq T \leq 3\times 10^8$~K. The lower boundary of the temperature range is set by our exclusion of molecular chemistry and detailed dust physics. The upper limit is arbitrary but is large enough to include many astrophysical regimes. Our module, therefore, can be applied from early phases of SN to ISM physics to ICM/IGM scales. 

The MHD equations for the density ($\rho$), velocity ($\vec{v}$), and magnetic field ($\vec{B} \equiv \vec{B}/\sqrt{4\pi}$ \footnote{the factor of  $1/\sqrt{4\pi}$ is absorbed in the definition of magnetic field in \pluto to avoid extra computation.}) in \pluto are written in conservative forms as
\begin{eqnarray}
&&\frac{\partial}{\partial t} \rho  + \vec{\nabla}\cdot\left(\rho \vec{v}\right) = \dot{\rho}_s \\
\label{eq:rho-cons}
&&\frac{\partial}{\partial t} \left(\rho \vec{v}\right) + \vec{\nabla}\cdot\left(\rho \vec{v} \otimes\vec{v} - \vec{B}\otimes\vec{B}+\overleftrightarrow{I}p_t \right)  = - \rho \vec{\nabla} \Phi + \rho \vec{a}_r \\
\label{eq:mom-cons}
&&\frac{\partial}{\partial t} \left(E+\rho\Phi\right) + \vec{\nabla}\cdot\left[\left(E + p_t + \rho\Phi\right)\vec{v} - \vec{B}\left(\vec{v}\cdot\vec{B}\right) \right] \nonumber \\
&& \quad\quad\quad\quad\quad= \mathcal{H}-\mathcal{L}  + \rho \vec{v}\cdot \vec{a}_r + \vec{\nabla}\cdot \vec{F}_c\\
\label{eq:energy-cons}
&& \frac{\partial}{\partial t}\vec{B}  - \vec{\nabla}\times\left(\vec{v}\times\vec{B}\right) = 0
\label{eq:induction}
\end{eqnarray} 
Here, $p_t = p + B^2/2$ is the total pressure (thermal + magnetic), $E = p/(\gamma-1) + \rho v^2/2 + B^2/2$ is the total energy and $\gamma=5/3$ is the adiabatic index. The source terms, $\dot{\rho}_s$, $\mathcal{H}$ and $\mathcal{L}$ are the mass injection rate, thermal heating and thermal cooling rates per unit volume, respectively. The thermal heating term usually includes  photo-heating and charge exchange heating, but can also include any external heating term. The radiation force and the conductive flux in Eqn \ref{eq:mom-cons}, \ref{eq:energy-cons} are given by $\rho\:\vec{a}_r$ and $\vec{F}_c$, respectively. All the source terms except $\dot{\rho}_s$ are solved using operator splitting.  The mass injection rate in the grid is only added as $\dot{\rho}_s\: dt$ after the end of each time step. This implementation requires that the mass be injected at zero velocity. All the details for solving the above equations can be found in \cite{Mignone2007} and \cite{Tesileanu2008} if not mentioned here.

Our module does not track the radiation energy density and, therefore, does not guarantee the conservation of radiation energy density in a Lagrangian element. This implies that the radiation can do mechanical work on the fluid, but the fluid does not do any mechanical work on the radiation. To overcome this problem, ideally, we would need to evolve the radiation energy density with time and treat it like a second fluid in the system. We reserve this issue for a future modification of code. Also, notice that we do not consider any magnetic field, and therefore any coupling of the magnetic field with the radiation is neglected in the tests mentioned here. 

\subsection{ionisation network}
\label{subsec:ion-network}
The ionisation network is solved by treating the ions as tracer particles inside the fluid but with a non-zero source function. The ion fraction $X_{k,i}$ of an ion $i$ of element $k$ is given by
\begin{equation}
\frac{\partial}{\partial t} X_{k,i} + \vec{v}\cdot \vec{\nabla}\: X_{k,i} = S_{k,i} 
\label{eq:ion-network}
\end{equation} 
where, $S_{k,i}$ contains the rate of ionisation and recombination of the ion ($k,i$) and is given as
\begin{eqnarray}
S_{k,i} &=&n_e \left[ X_{k, i+1}\alpha_{k, i+1} - X_{k,i} \left( \xi_{k,i} + \alpha_{k,i}\right) + X_{k, i-1} \xi_{k,i-1} \right] \nonumber \\
  &-& X_{k,i} \Gamma_{k,i} + A_{k,i}
\end{eqnarray}
Here, $n_e$ is the electron density, $\alpha_{k,i}$ is the total recombination rate for ion $(k,i)$ to $(k, i-1)$ and $\xi_{k,i}$ is the total ionisation rate of ion $(k,i)$ to $(k, i+1)$, $\Gamma_{k,i}$ is the photo-ionisation rate and $A_{k,i}$ is the Auger ionisation rate of lower ions to the current ion.

 The ionisation and recombination rates also include charge transfer (CT) rates and are given in detail as 
\begin{eqnarray}
\xi_{k,i} &=& \xi_{k,i}^{\rm coll} + \frac{n_{\rm HII}}{n_e} \xi_{k,i}^{\rm HII} + \frac{n_{\rm HeII}}{n_e} \xi_{k,i}^{\rm HeII} \nonumber \\
\alpha_{k,i} &=& \alpha_{k,i}^{\rm diel+rad} + \frac{n_{\rm HI}}{n_e} \alpha_{k,i}^{\rm HI} + \frac{n_{\rm HeI}}{n_e} \alpha_{k,i}^{\rm HeI}
\end{eqnarray}
The total ionisation rate ($\xi_{k,i}$) thus consists of (from left to right) the collisional ionisation rate and the CT rates with HII and  HeII. The total recombination rate ($\alpha_{k,i}$) includes the radiative+dielectronic recombination rates and the CT rates with HI and HeI. The CT reactions of the metals with H and He could also be included in the rate equations of H and He but since the number density of H and He are overwhelmingly large compared to  metals, this rate does not affect the H and He ion fractions. 
We use the rate coefficients used in \cite{Gnat2007}. In addition, we also include the statistical CT rates for ions with charge $\geq 4+$ as prescribed by \cite{Ferland1997}\footnote{The statistical CT is assumed only for the highly ionised elements considering that such highly ionised ions have so many energy levels available that a mere collision with a neutral atom can cause charge transfer. }

In our new module we also include the photo-ionisation effects. The photo-ionisation rate for a given spectrum $J_\nu$ ($\ergpspcmsq$ sr$^{-1}$) is
\begin{equation}
\Gamma_{k,i}= \sum_{s = 1s}^{s \leq 4s} \int_{\nu_{\rm IP, s, k,i}}^{\infty} \frac{4\pi J_\nu}{h\nu} \sigma_{\nu,k,i,s}^{\rm pi} d\nu  
\end{equation} 
where, $\sigma_{\nu,k,i,s}$ is the photo-ionisation cross section of the ion at a given frequency. Here, $s$ represents the shell numbers ($1s, 2s, 2p, ...., 4s$) of the remaining electrons. The sum over the shells is needed to calculate the total photo-ionisation cross section for the ion as any of the electrons from an inner shell can also be ejected by this process. The sum does not include electrons higher than $4s$ which is required only in atoms with atomic number $>30$. The cross sections are obtained from the fits provided in \cite{Verner1996a}.
We include stripping of inner electrons by Auger processes. The Auger rate can be written as 
\begin{equation}
A_{k,i} = X_{k,i} \sum_{g<i} \:\sum_{s} P_{k, g, s}(i-g) \int_{\nu} \frac{4\pi J_{\nu}}{h\nu} \sigma_{\nu,k,g,s} d\nu \ \ \ ,
\end{equation} 
where $P(N)$ is the probability for ejecting $N$ electrons.

Eq \ref{eq:ion-network} for each ion is solved using a split source method, i.e. the right-hand side is assumed to be zero while advecting the ions and considered only afterwards for obtaining the temporal evolution. The numerical integration uses Embedded Runge-Kutta (RK) methods such as Runge–Kutta–Fehlberg, if the reaction rate is not very high, i.e. not `stiff', or  Cash-Karp if the equations are `stiff'. For stiff conditions, the hydrodynamic timescale is further divided into sub-steps to reduce the error on the RK method. Full details of above methods can be found in \cite{Tesileanu2008}.
\subsection{Cooling and Heating}
\label{subsec:cooling-heating} 
The radiative cooling term, $\mathcal{L}$ ($\ergps \pcc$), includes recombination, free-free, and collisionally excited line radiation terms. The total radiation efficiency, $\Lambda_{k,i}$ ($\ergps \cc$), for each ion is taken from pre-computed tables from \cloudy-17 and similar to the ones given in \cite{Gnat2012}. The total cooling rate for all the ions is given in terms of the electron density, $n_e$, and ion density, $n_{k,i}$ as
\begin{equation}
\mathcal{L} =  n_e \sum_{k,i} n_{k,i}\:\Lambda_{k,i}(T) 
\label{eq:cooling}
\end{equation} 
Also, the cooling rates can be easily updated with a newer version of \cloudy. 
We stress that this implementation assumes an 
coronal level population configuration of electrons for each ion. 

Our code also works 
for a zero metallicity case with the only H and He included in the NEI network as well as with a trimmed network for the metals. For the trimmed network, the sum of the ion fractions is kept equal to unity for a given element. Since the trimmed network can significantly change the ion fraction of the highest-NEI level, the cooling functions may be affected. To remedy this, we modify the cooling efficiency of the highest-NEI level by weighing it against the ion fraction of the higher level in CIE (following \citealt{Gray2015}). 
\begin{equation}
\Lambda_{i_{\rm max}}(T) = \frac{\sum_{i\geq i_{\rm max}} X_{i,\mbox{cie}}(T)\, \Lambda_{i, \mbox{cie}}(T) }{\sum_{i\geq i_{\rm max}} X_{i,\mbox{cie}}(T)} 
\end{equation}
where $X_{i, \mbox{cie}}$ and $\Lambda_{i,\mbox{cie}}$ are the ion fractions and cooling functions of the ions in CIE. 
This is a valid assumption when the maximal level in the trimmed network is not expected to significantly depart from equilibrium (for example, ions with charge $\gtrsim 4+$).
For lower ions (say, charge $= +1$), this approximation fails completely. We do not use (or recommend using) such a highly trimmed network.

The heating is taken as 
\begin{eqnarray}
\mathcal{H} &=& \sum_{k, i, s} n_{k,i} \int_{\nu_{\rm IP, s, k,i}}^{\infty}   \frac{4\pi J_\nu }{h\nu} \left(h\nu-h\nu_{\rm IP, s, k,i} \right) \sigma_{\nu,k,i,s}^{\rm pi} d\nu  \nonumber \\
&+& \sum_{k, i} n_{k,i} \left( n_{\rm HI}\: \alpha_{k,i}^{\rm HI}\: \Delta_{k, i}^{\rm rec} +  n_{\rm HII}\: \xi_{k,i}^{\rm HII}\: \Delta_{k, i}^{\rm ion}\right)
\label{eq:heating}
\end{eqnarray}
 Here, the first term is the usual photo-heating term and the second term is the CT heating/cooling term and only considered if charge is transferred with H ions. The recombination or ionisation energy, $\Delta_{k, i}$, for each CT case is taken from \cite{Kingdon1999} following \cloudy-17.
 
\subsection{Conduction}
\label{subsec:conduction}
We use the pre-existing conduction module in \pluto.  The conductive flux is given as 
\begin{equation}
\vec{F}_c = \frac{F_{\rm sat}}{F_{\rm sat} + |\vec{F}_{\rm class}|} \vec{F}_{\rm class}\,,
\end{equation}
 where, $\vec{F}_{\rm class}$ is the classical Spitzer conductive flux in the absence of magnetic field and $F_{\rm sat} = 5 \phi\:\rho\:c_{\rm iso}^3$ is the saturated flux when the temperature gradient scale is smaller than the electron mean free path. 
 When using thermal conduction (so far isotropic), we set $F_{\rm class} = 5.6\times 10^{-7}\: T^{5/2}\: \nabla T\: \ergpspcmsq $ following \cite{Spitzer1956} and $\phi = 0.3$ following \cite{Cowie1977}.
 
\subsection{Radiative transfer}
\label{subsec:rad-trans}
Frequency dependent radiative transfer (RT) is solved at the beginning of each source-splitting loop assuming the light crossing time is much shorter than the typical time-scale for the hydrodynamics to change. The RT equation in a spherically symmetric system is then
\begin{equation}
\frac{\mu}{r^2}\, \frac{\partial}{\partial r}\left( r^2\, \psi(\mu, r,\nu)\right) + \frac{1}{r} \frac{\partial}{\partial \mu}\left( (1 - \mu^2) \psi(\mu, r,\nu)\right) = j_\nu - \alpha_\nu\, \psi(\mu, r,\nu) \,,
\label{eq:RTE-sph}
\end{equation}
 where, $r$ is the radius, $\mu = \cos\theta$, is the cosine of the angle subtended by a ray with the radial direction,  $\psi$ ($\ergpspcmsq$ Hz$^{-1}$ sr$^{-1}$)  is the specific intensity of a ray, $\alpha$ (cm$^{-1}$)  is the absorption coefficient (we will loosely 
 refer to it as opacity throughout the text) and $j_\nu$ ($\ergps \pcc$ Hz$^{-1}$ sr$^{-1}$) is the emissivity. Notice that we have removed the derivative with respect to the frequency $\nu$ in the above equation which assumes that all the velocities are non-relativistic so that no energy is transferred across frequency bands. This assumption is particularly justified if the frequency band width ($\Delta \nu$) at any frequency ($\nu$) is such that $\Delta \nu /\nu \ll v/c$, where $v/c$ is the velocity of a fluid element compared to speed of light. Our RT equation is only applicable in systems where scattering in the given frequency range is negligible and $\alpha$ is purely dominated by absorption.  Our equation demands that the emission and absorption coefficients are spherically symmetric and $\psi$ is axisymmetric about the radial direction.
 
 The absorption coefficient is calculated at each time step at each frequency band as 
 \begin{equation}
 \alpha_\nu = \sum_{k,i} n_{k,i} \sum_{s} \sigma_{k,i,s,\nu}^{\rm pi}\,.
 \end{equation}
For $j_\nu$, we assume an isotropic collisional equilibrium emissivity which only depends on the total hydrogen density, $n_H$ and temperature $T$. This emissivity is obtained from \cloudy-17 for a given metallicity of $Z_\odot$ without the presence of any metagalactic radiation field.
Thus, while our IN and RT are consistent with each other, the assumption that $j_\nu(T)$ is an equilibrium emissivity is not fully consistent with the local non-equilibrium ion-fractions. 
The computation of the non-equilibrium emissivity could be done via iteration since the ion-fractions also partially depend on the emissivity.

\subsection{Dust}
Dust plays an important role in the interstellar medium, contributing to extinction, and mediating radiation pressure and thermal heating
\citep{Trumpler1930, Dopita2000, Draine2011}. Although dust absorption or radiation pressure provided by dust is not very significant in a low density medium, it can play a major role  
at higher densities. For example, for a \strom sphere, we can compare the \strom radius ($R_{st}$) and the mean free path ($\lambda_d = 1/ n_H \sigma_{\rm ext, d}$) for a photon in a dusty medium, where $\sigma_{\rm ext, d}$ is the dust extinction cross section per H nuclei. This produces a lower limit to the density above which the dust becomes important
\begin{equation}
n_{H, dcrit} \simeq 100\: Q_{49}^{-1}\: T_4^{-0.84}\: \sigma_{ext, d, -21}^{-3}
\label{eq:nH-dcrit}
\end{equation}
where, $ \sigma_{ext, d, -21} =  \sigma_{ext, d}/10^{-21}$ cm$^2$.
This means that in molecular clouds with densities $\gtrsim 100 \pcc$, extinction and radiation pressure offered by dust will be very important in the ionisation front dynamics \citep{Spitzer1998}. 

To make our code suitable for studies at higher densities, we include a very simple prescription for dust extinction. We consider extinction tables provided by \cite{Weingartner2001} for $R_V = 3.1$ which is very close to the observed dust properties in the Milky-Way\footnote{Available in \url{https://www.astro.princeton.edu/~draine/dust/dustmix.html}}.  This table provides the extinction cross section per H nuclei  ($[\sigma_{\rm ext, d} ]$), the albedo ($\omega$) and the average angle of scattering ($<\cos \theta>$) as a function of frequency. The total extinction, scattering and absorption opacities are, therefore, given as 
\begin{eqnarray}
\alpha_{\rm ext,d} &=& n_H\: \sigma_{\rm ext, d} \nonumber \\
\alpha_{\rm scat,d} &=& \omega\: \alpha_{\rm ext,d} \nonumber \\
\alpha_{\rm abs,d} &=& \alpha_{\rm ext,d} - \alpha_{\rm scat,d} \,.
\end{eqnarray}
Now since the scattering can happen at any direction, the opacity to be used for the radiation pressure is not the same as the total extinction opacity. It is given by
\begin{equation}
\alpha_{\rm pr,d} = \alpha_{\rm abs,d} + (1 - <\cos\theta>)\: \alpha_{\rm scat,d}.
\end{equation}
The total opacity (gas + dust) to be used while solving the radiative transfer is therefore $\alpha = \alpha_{\rm gas} + \alpha_{\rm ext,d}$, and for the radiation acceleration $\alpha_{\rm gas} + \alpha_{\rm pr,d}$. We do not  model the scattered light from the dust as it is mostly in the infrared, which is outside our considered frequency range. We also assume that the photo-electron heating from the dust is negligible compared to photo-ionisation heating from the gas and we do not consider dust heating in our code.

Since dust can be easily destroyed in shocks or in a hot medium, we include a simple prescription for dust sputtering from \cite[][eq 25.13]{Draine2011}. This rate is given as 
\begin{equation}
\frac{d\:a}{dt} = -\frac{10^{-6}}{1+T_6^{-3}} \, n_H \qquad \mu\mbox{m yr}^{-1}.
\end{equation}
Although the extinction curve used in this work consists of a mixture of different dust particle sizes, we assume that the dust sputtering is well represented by a single population of dust particles with the initial size of $a = 0.1 \mu$m  (approximately the wavelength for a $10$ eV photon). The actual dust opacity at any later time is then multiplied by $\left( a(t)/0.1\mu m \right)^2$ to account for the dust destruction.  
We emphasise that this is a `proof of principle' aimed at implementing dust opacity rather than an attempt to model a `true' physical description of dust and its interaction with the ISM.  We keep these modifications for a future upgrade of the code. 
\section{Numerical technique to solve Radiative Transfer}
\label{sec:num-rt-technique}
We use the method of short characteristics to solve the RT equation in spherical symmetry. 
In many applications, the radiative transfer is important mainly for its influence on dynamics
rather than the exact effects on the ionisation of atoms. Our focus is on the ionisation states themselves.
 In addition,  we want accurate solutions even at the transition layers between optically thick and thin regions since these may host rarer ions and reactions.  This is why we employ the method of short characteristics \citep{Lewis1984}.
 
We solve eq \ref{eq:RTE-sph} by discretising the $r-\mu$ space in spatial and angular grids represented by $r_i$ and $\mu_m$, respectively. The spatial discretisation is the same as used in solving the hydrodynamics. The angular coordinate is discretised in uniform grids between $\mu = -1$ to $+1$. Such a uniform  discretisation
allows us to avoid the issues encountered in the free streaming limit when the energy preferably flows along $\mu= \pm 1$.  While writing the discretised form of eq \ref{eq:RTE-sph}, we use a `finite volume' method which is more accurate in conserving the flow of energy between angular/radial bins than the `finite element' method for a given number of bins. To obtain a finite volume-like form, we integrate the equation between two grid points. We assume that $j_\nu$ and $\alpha_\nu$ are isotropic and remain constant within a radial cell, $\psi$, however, varies linearly in both $r$ and $\mu$ within a single cell. Similar methods have been used previously in different systems, from neutron diffusion in nuclear reactors \cite{Hill1975} to neutrino transport inside supernovae \cite{Birnboim2000}. These works, however, use a diamond difference
scheme (constant $\psi$ between radial/angular cells) which is $\mathcal{O}(h^2)$ accurate in estimating $\psi$ at the cell edges, where $h = \alpha_\nu (r_{i+1}-r_{i-1})/ (2 |\mu_m|)$. We assume a linear variation of $\psi$ in both $r$ and $\mu$ directions. This is expected to increase the accuracy of the method to $\mathcal{O}(h^3)$ suitable for rays where either optical depth across a cell is high, or there is substantial angular variation in intensity \citep{Larsen1982}. 
We also use a better technique for fixing negative intensities, as we shall explain.

Integrating the above equation first in the range from $\mu_m$ to $\mu_{m+1}$ and then from $r_i$ to $r_{i+1}$  produces (removing explicit $\nu$ dependence for convenience)\footnote{For more details, see Appendix \ref{app-sec:rad-trans-method}.}
\begin{equation}
a_{i,m} \:\psi_{i,m} + b_{i,m}\:\psi_{i+1,m} + d_{i,m}\:\psi_{i,m+1} + f_{i,m}\:\psi_{i+1,m+1} = j_i\:\Delta\mu_m\: \frac{\Delta V_i}{4\pi}
\label{eq:RTE-matrix-eq}
\end{equation}
where, $\Delta V_i = \frac{4\pi}{3} (r_{i+1}^3-r_i^3)\:$, $\:\Delta \mu_m = \mu_{m+1}-\mu_{m}$ and the time varying coefficients are given as 
\begin{eqnarray}
a_{i,m} &=& -A_m r_i^2 -(1-\mu_m^2)\: C_i + \frac{\alpha_i\:\Delta\mu_m}{2} B_i \nonumber \\
b_{i,m} &=& A_m r_{i+1}^2 - (1-\mu_m^2)\:\bar{C}_i + \frac{\alpha_i\:\Delta\mu_m}{2} \bar{B}_i  \nonumber \\
d_{i,m} &=&  \bar{A}_m r_i^2 + (1-\mu_{m+1}^2)\: C_i + \frac{\alpha_i\:\Delta\mu_m}{2} B_i \nonumber \\
f_{i,m} &=&  -\bar{A}_m r_{i+1}^2 + (1-\mu_{m+1}^2)\: \bar{C}_i + \frac{\alpha_i\:\Delta\mu_m}{2} \bar{B}_i \,.
\label{eq:matrix-coeffs}
\end{eqnarray}
The constant coefficients are
\begin{eqnarray}
A_m &=& \frac{\Delta\mu_m}{6}\:(\mu_{m+1}+2\mu_m) \nonumber \\
\bar{A}_m &=&  -\frac{\Delta\mu_m}{6}\:(2 \mu_{m+1}+\mu_m) \nonumber \\
B_i &=& \frac{1}{12\: \Delta r_i}\: \left( r_{i+1}^4 - 4\:r_i^3\:r_{i+1} + 3\:r_i^4\right) \nonumber \\
\bar{B}_i &=& \frac{1}{12\: \Delta r_i}\: \left(3 r_{i+1}^4 - 4\:r_i\:r_{i+1}^3 + \:r_i^4\right) \nonumber \\
C_i &=& \frac{\Delta r_i}{6} \left(r_{i+1}+ 2\:r_i \right)  \nonumber \\
\bar{C}_i &=& \frac{\Delta r_i}{6} \left(2\:r_{i+1}+r_i \right)\,.
\label{eq:ABC}
\end{eqnarray}

For a given emissivity, eq \ref{eq:RTE-matrix-eq} can be solved by 
inverting a $(N_r N_\mu)\times (N_r N_\mu)$ matrix, but this is time consuming for reasonable numbers of the $r, \mu$ grids. We rather follow a different approach which uses a special analytical solution of equation \ref{eq:RTE-sph} along $\mu = -1$ and spherical symmetry at the innermost boundary in $r$. 

We choose our $\mu$-grids to be symmetric around $\mu = 0$, i.e. $\mu_m = -1, \mu_1, \mu_2, ...., \mu_{N_\mu/2-1}, \mu_{N_\mu/2}, ...., +1$ where the values $\mu_0- \mu_{N_{\mu/2}}$ have negative values but $\mu_{N_{\mu/2}+1}$ and onward have positive values. We start from the outer boundary, at $i = N_r-1$ and $m=\mu_0$ to $\mu_{N_\mu}/2$, and specify the background radiation field irradiated on the system as the outer boundary condition. 
 We then solve the RT along $\mu = -1 $ where the solution is not dependent on the angular derivative and is given simply as 
\begin{equation}
\psi_i (-1) = \psi_{i+1} (-1)\,\exp\left(- \alpha_i\: \Delta r_i \right)+ \frac{j_i}{\alpha_i}\, \left[1-  \exp\left(- \alpha_i\: \Delta r_i \right)  \right]  
\label{eq:central-ray}
\end{equation} 
\begin{figure}
	\centering
	\includegraphics[width=0.5\textwidth, clip=true, trim={2cm 1cm 1cm 1cm}]{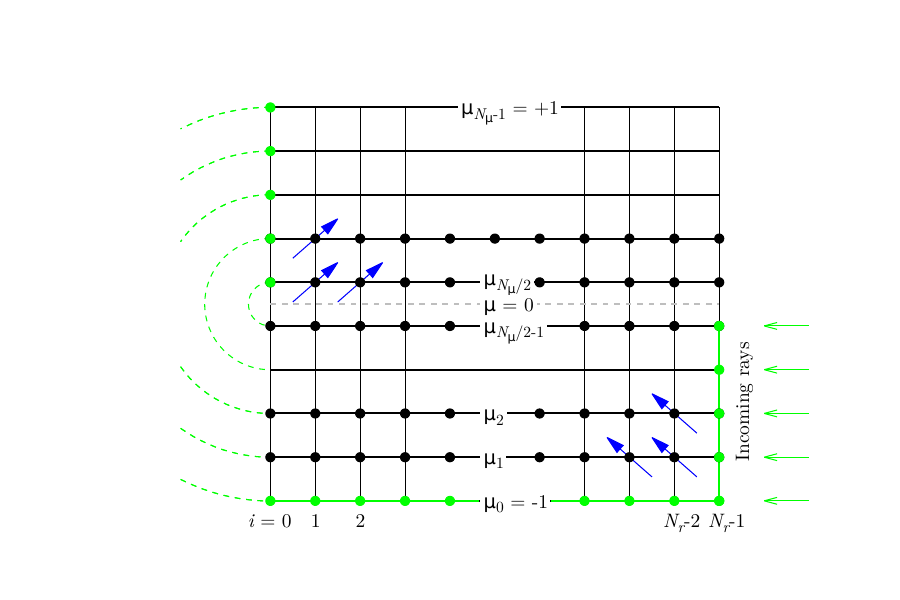}
	\caption{$r - \mu$ grid to solve radiative transfer equation \ref{eq:RTE-matrix-eq}. The green solid line at $i = N_r-1$ represents the incoming rays (the  background), the green line along $\mu = -1$ represents the radially incoming ray for which we have obtained an analytical solution (Eq \ref{eq:central-ray}). The green filled circles at $i = 0$ and $n = N_\mu/2$ to $ N_\mu-1$ represent the grids at the inner boundary where the spherically symmetric assumption has been applied to copy the values from $\mu<0$ rays as indicated by the dashed green lines. The propagation of information in the $\mu<0$ and $\mu>0$ region is shown by the blue arrows. }
	\label{fig:r-mu-grid}
\end{figure} 

The grid to solve eq \ref{eq:RTE-matrix-eq} is shown in figure \ref{fig:r-mu-grid}. The obtained boundary conditions are shown by the green lines and dots.  Given this boundary condition and eq \ref{eq:RTE-matrix-eq}, we can write down 
\begin{equation}
\psi_{i,m+1}  = \frac{1}{d_{i,m}}\left[ j_i\:\frac{\Delta\mu_m \Delta V_i}{4\pi} - ( a_{i,m}\psi_{i,m} + b_{i,m}\psi_{i+1,m} + f_{i,m}\psi_{i+1,m+1})\right] \,\,,
\label{eq:soln-incoming-rays}
\end{equation}
which means that given $\psi_{N_r-2,0}\,, \psi_{N_r-1,0}$ and $\psi_{N_r-1,1}$ we can determine the value of $\psi_{N_r-2,1}$. This procedure can be applied to obtain $\psi_{i,m}$ for all $i=0-(N_r-1)$ and $m=0 - (N_\mu/2-1)$. However, it cannot be extended for $m = N_\mu/2$ to $+1$ as we do not have the prior information of the outgoing ray ($\mu>0$) at the outer boundary. Fortunately, we can apply the spherically symmetric condition at the inner boundary of the sphere, i.e. $\psi_{0, N_\mu-1} = \psi_{0, 0}\,\,\,,\,\, \psi_{0, N_\mu-2} = \psi_{0, 1}$ and so on.  This allows us to write (for the $\mu>0$ region)
\begin{equation}
\psi_{i+1,m+1}  = \frac{1}{f_{i,m}}\:\left[ j\:\Delta\mu\: \frac{\Delta V}{4\pi} - \left( a_{i,m} \:\psi_{i,m} + b_{i,m}\:\psi_{i+1,m} + d_{i,m}\:\psi_{i,m+1} \right)\right] \,\,,
\label{eq:soln-outgoing-rays}
\end{equation}
which means that given $\psi_{0,N_\mu/2-1}\,, \psi_{1,N_\mu/2-1}$ and $\psi_{0,N_\mu/2}$ we can find out the value of $\psi_{1, N_\mu/2}$. This method, as before, then can be applied to the rest of the grid. Notice that the propagation of information in this way of solving for the $\psi$ follows the overall direction of photon travel and, therefore, increases the stability of the algorithm \citep{Lewis1984}. Now, once $\psi$ for all the grids have been calculated, we can find the angular averaged intensity and radiative flux (see section \ref{subsec:intensity+flux} and eq \ref{eq:J-F}).

\begin{figure}
	\centering
	\includegraphics[width=0.35\textheight, clip=true, trim={1cm 19.5cm 2cm 1.5cm}]{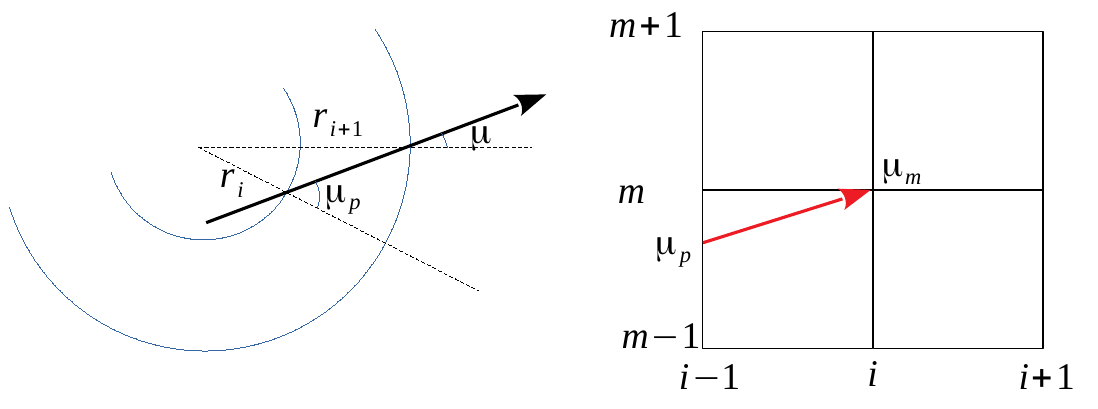}
	\caption{Fixing negative intensity by \textit{truestream}. The left panel shows the physical geometry, whereas the right panel shows the $r-\mu$ grid structure. An outgoing ray, as an example, has been shown by the solid arrow. }
	\label{fig:intensity-fix-geometry}
\end{figure}
\subsection{Fixing negative intensity}
\label{subsec:fix-neg-intensity}
An important issue with the above method is that the assumed linear approximation of  $\psi$ in $r-\mu$ grid can break down (since the accuracy is only $\mathcal{O}(h^3)$), for example, when a ray peaks very sharply only along a single direction, say, $\delta(\mu)$. In such cases, the linear interpolation predicts excess energy flow from a cell to its neighbouring cell which results in an overall negative intensity from the cell. In such cells we no longer accept values given by eq \ref{eq:RTE-matrix-eq}, rather use another method to obtain the solution. We refer this as the \textit{truestream} method. In this method, if a grid point  $(r_i,\mu_m)$  faces a negative intensity, we track individual rays from the previous grid to the current grid. This is done in two steps. First, we find the origin of the given ray at the previous grid ($r_{i-1}$ for $\mu_m>0$ sweep, for example), say $\mu_p$. This value is given as 
\begin{eqnarray}
\mu_p &=& \sqrt{1-\left(\frac{r_i}{r_{i+1}}\right)^2 \left( 1-\mu_m^2 \right) } \quad\quad\quad \mbox{for } \mu_m < 0 \nonumber \\
 &=& \sqrt{1-\left(\frac{r_i}{r_{i-1}}\right)^2 \left( 1-\mu_m^2 \right) } \quad\quad\quad \mbox{for } \mu_m > 0
\end{eqnarray}
The intensity at this angle is then found by simple linear interpolation between the two adjacent grids, $r_{i-1},\mu_{l}$ and $r_{i-1},\mu_{h}$ (see figure \ref{fig:intensity-fix-geometry}) as
\begin{equation}
\psi_p = \frac{1}{d\mu_l} \left[ (\mu_p - \mu_l) \psi_{i-1,h} + (\mu_h - \mu_p) \psi_{i-1,l} \right]
\end{equation}
Note that $\mu_l$ and $\mu_h$ can be anywhere along the ray vector $\mu_m$ and have to be searched for. Now, once we have found $\psi_p$ at $r_{i-1}$, we can find $\psi_{i,n}$ as
\begin{equation}
\psi_{i,m} = \psi_p \exp\left( - \alpha_{i-1} x\right) + \frac{\epsilon_{i-1}}{\alpha_{i-1}}\,\left[ 1-  \exp\left( - \alpha_{i-1} x\right)\right]
\end{equation}
where, 
\begin{eqnarray}
x &=& \frac{\sin(\theta - \theta_p)}{\sin\theta_p}\: r_i \quad\quad\quad \mbox{for } \mu_m < 0 \nonumber \\
 &=& \frac{\sin(\theta_p - \theta)}{\sin\theta_p}\: r_i \quad\quad\quad \mbox{for } \mu_m > 0
\end{eqnarray}
with $\theta = \cos^{-1}(\mu_m)$ and $\theta_p = \cos^{-1}(\mu_p)$. The advantage of this method is that we recalculate the intensity of that grid in an exact way.  Unconditionally applying this method throughout the grid can result in slow down of the code. 

Sometimes, the above method can ignore the rays that do not pass through other cells but only migrates from $\mu<0$ half to $\mu>0$. Such cases may arise at the very inner radii where $\mu_m < \sqrt{1-(r_{i-1}/r_i)^2}$ for a ray. 
In such cases $\mu_m = -\mu_m$ and, therefore, 
\begin{eqnarray}
\psi (i,m) &=& \psi(i,N_\mu-m-1)\: e^{ -2 r_i\: \mu_m\: \alpha_{i-1}} \nonumber \\
   &+& \frac{j_{i-1}}{\alpha_{i-1}} \left(1- e^{-2 r_i\: \mu_m\: \alpha_{i-1}} \right) 
\end{eqnarray}
where, $2 r_i\mu_m$ is simply the path length travelled by the ray in that given cell.
\begin{figure*}
 \centering 
 \includegraphics[width=0.4\textwidth, height=0.9\textheight,  clip=true, trim={0.7cm 3cm 0.8cm 2.2cm}]{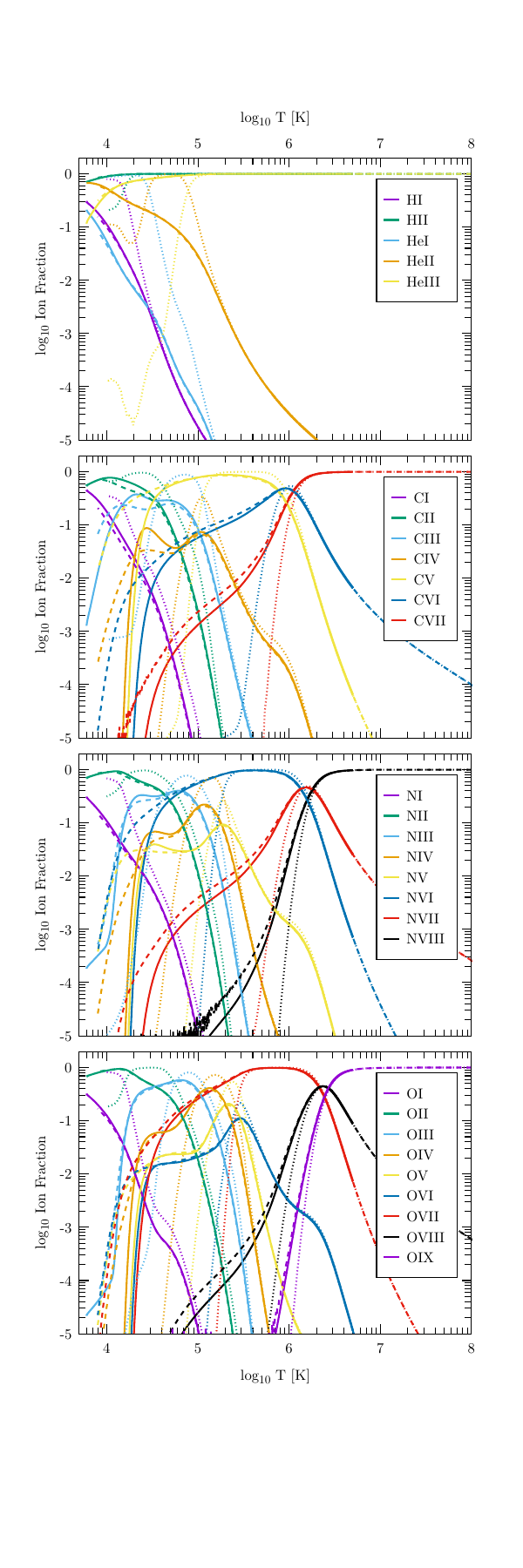}
 \includegraphics[width=0.4\textwidth, height=0.9\textheight,  clip=true, trim={0.7cm 3cm 0.8cm 2.2cm}]{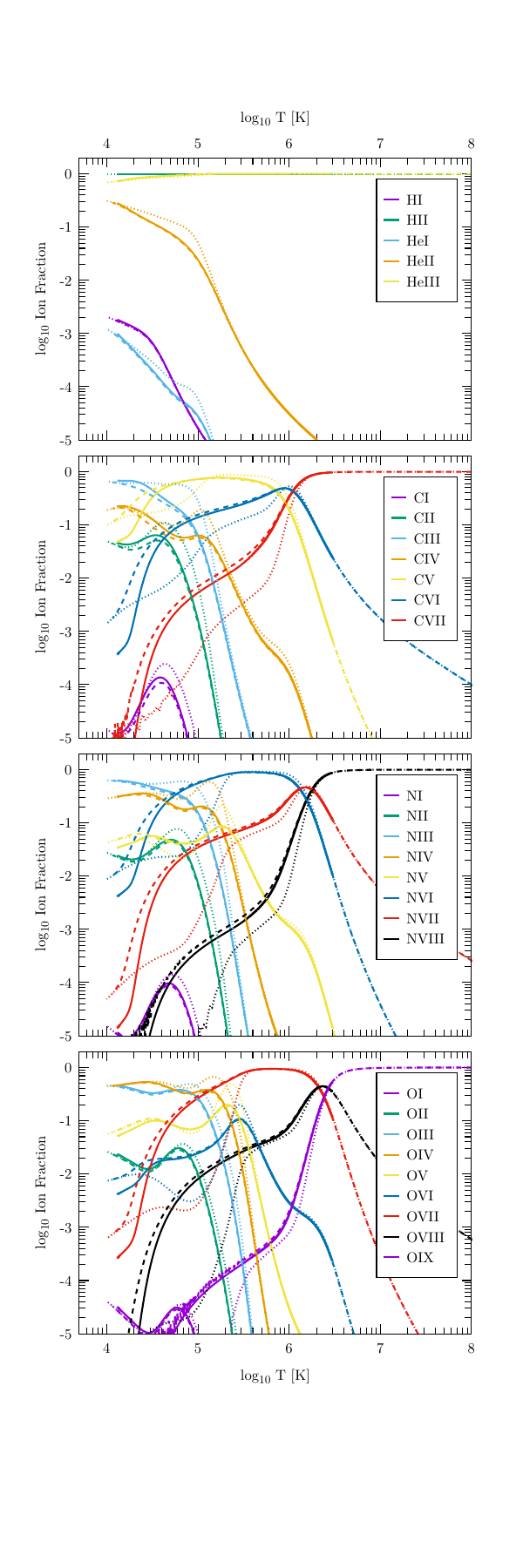}
   \caption{ Evolution of H, He, C, N and O ions under isochoric cooling for pure photo-equilibrium (dotted lines), G17 results (dashed lines) and new results (solid lines) for isochorically cooling gas. \textit{Left column:} for $n_H = 1 \pcc$ and \textit{Right column:} for $n_H = 10^{-4} \pcc$. Notice how NEq evolution affects the over-ionisation of certain ions. The difference between G17 and the new results are due to the inclusion of statistical charge transfer for higher ions.}
	\label{fig:isochoric-ions}
\end{figure*}

\subsection{Intensity and Flux}
\label{subsec:intensity+flux}
The angular averaged specific intensity (or the mean intensity), and total flux (radially outwards) at any $r_i$ and $\nu$ can be written as 
\begin{eqnarray}
J_i &=& \frac{1}{2}\: \int_{-1}^{1} \psi\: d\mu \,=\, \sum_{m=0}^{N_\mu}\frac{\Delta\mu_m}{4}\:\left(\psi_m + \psi_{m+1} \right) \quad \mbox{   and} \nonumber \\
F_i &=&  \int_{0}^{2\pi} d\phi \:\int_{-1}^{1} \psi\:\mu\: d\mu\,=\,2 \pi\:\int_{-1}^{1} \psi\:\mu\: d\mu \nonumber \\
&=& 2 \pi\:\sum_{m=0}^{N_\mu}\left( A_m \psi_m - \bar{A}_m\psi_{m+1} \right)
\label{eq:J-F}
\end{eqnarray}
following the same assumption of linear interpolation as before. However, notice that these quantities are, by construction, face centred unlike the cell centred hydrodynamic quantities. We, therefore, use the volume averaged values of the $J$ and $F$ at that cell. It is easy to see from Eq \ref{eq:ABC} that for any quantity that is assumed to vary linearly within a cell $r_{i}$ to $r_{i+1}$, the volume averaged values are given by 
\begin{eqnarray}
J &=& \frac{3}{r_{i+1}^3-r_i^3}\, \left( B_i\: J_i + \bar{B}_i\: J_{i+1}\right) \nonumber \\
F &=& \frac{3}{r_{i+1}^3-r_i^3}\, \left( B_i\: F_i + \bar{B}_i\: F_{i+1}\right)\,\,.
\end{eqnarray}
We use these volume averaged values for the calculation of photo-ionisation rates and radiative force on any cell.

%
\section{Tests}
In this section, first, we show that our ionisation network (IN) and radiative transfer (RT) procedure do work separately and then show how they work together. In all the tests, we used a Solar metallicity as given in \cite{Asplund2009}. 
 \begin{figure*}
	\includegraphics[width=0.35\textheight, clip=true, trim={0cm 0cm 0cm 0cm}]{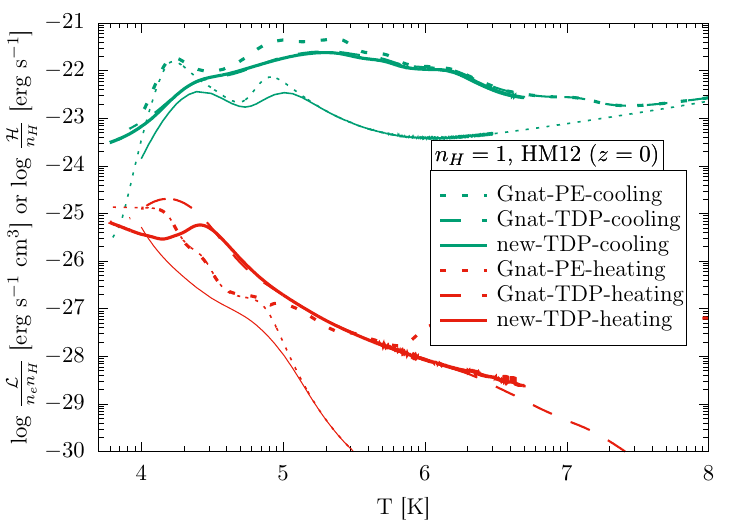}
	\includegraphics[width=0.35\textheight, clip=true, trim={0cm 0cm 0cm 0cm}]{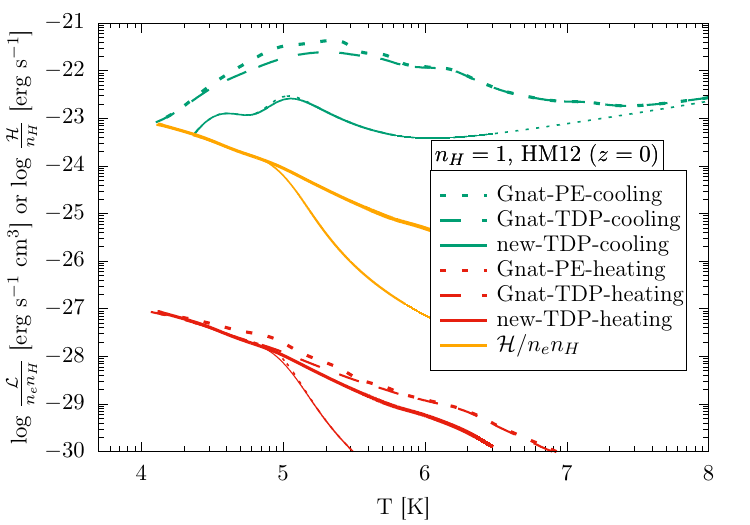}
	\caption{Heating and cooling functions for isochoric cooling in the presence of HM12 ($z = 0$) radiation. \textit{Left panel}: for $n_H = 1\pcc$ and \textit{right panel} is for $n_H = 10^{-4} \pcc$. The dotted lines show the cooling/heating functions in the case of a photo+collisional equilibrium (PE) case, the dashed lines represent a time-dependent isochoric cooling (TDP) case from G17, and the solid lines show the new results. The orange line in the right panel shows the heating, $\mathcal{H}/n_e\:n_H$ ($\ergps \cc$), for our new results to compare with the cooling. The thinner dashed and solid lines represent the corresponding heating and cooling for plasma with zero metallicity.}
	\label{fig:isochoric-heating-cooling}
\end{figure*}

\begin{figure*}
	\includegraphics[width=0.35\textheight, clip=true, trim={0cm 0.5cm 0.5cm 0cm}]{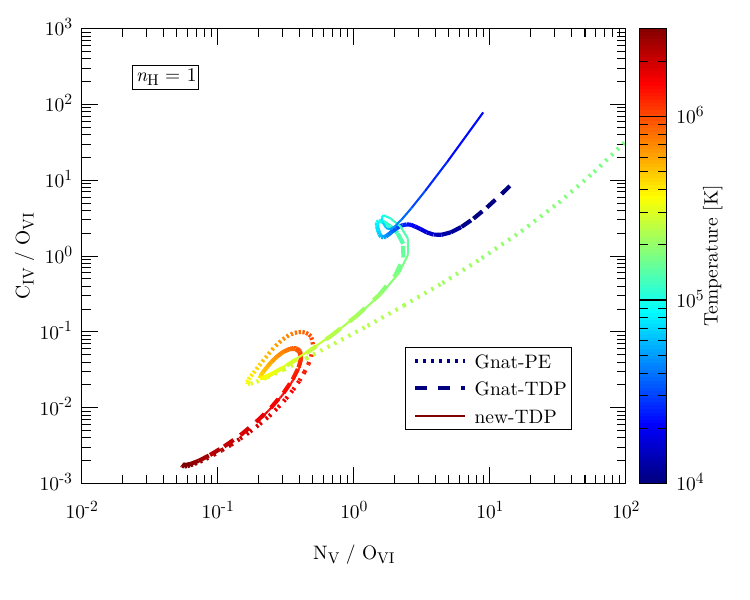}
	\includegraphics[width=0.35\textheight, clip=true, trim={0cm 0.5cm 0.5cm 0cm}]{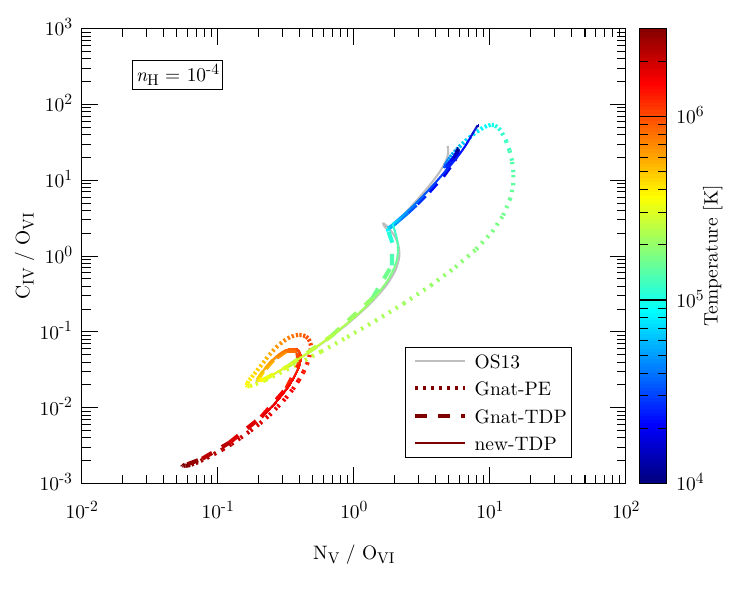}
	\caption{ \ion{N}{v}/\ion{O}{vi} vs \ion{C}{iv}/\ion{O}{vi} as a function of temperature of the plasma (shown in colour palette). The dotted lines show a photo+collisional equilibrium case, the thick dashed line shows the results from G17 and the thin solid line shows the new results. \textit{Left panel} is for $n_H = 1\pcc$ and \textit{Right panel} is for $n_H = 10^{-4} \pcc$. The ratio evolves as the plasma cools to lower temperatures as indicated by the colour palette. 
	A comparison with the results from \protect\cite{Oppenheimer2013} is shown by the solid grey line (hardly distinguishable from our computation) in the right panel.}
	\label{fig:isochoric-ion-ratio}
\end{figure*}
\subsection{Ionisation network}
\label{subsec:tests-ion-net}
We test the IN by following a zero-dimensional simulation where an initial hot plasma ($T = 5\times10^6$~K) is allowed to cool isochorically to a floor temperature ($T = 5 \times 10^3$~K) in the presence of a metagalactic radiation field taken from \cite{Haardt2012} (hereafter, HM12) at redshift zero ($z = 0$). This particular kind of test has been performed several times in the literature, as mentioned earlier. The results, however, are somewhat dependent on the atomic data used. In this particular test, we compare our results with \cite{Gnat2017} (hereafter, G17).

Figure \ref{fig:isochoric-ions} shows the evolution of the ion fractions for non-equilibrium isochoric cooling using our code (by solid) and from G17 (dashed lines). Equilibrium ion fractions in the presence of the same metagalactic radiation are shown by dotted lines for reference.  At high temperatures ($T \gtrsim 10^6$~K), the cooling time is much longer than the ionisation or recombination time. Therefore the ions remain in equilibrium. The situation changes at lower temperatures ($T \lesssim 3 \times 10^5$~K) when $\tau_{\rm cool} < \tau_{\rm rec}$ and the gas 
departs from equilibrium.
Such non-equilibrium effects are much more prominent at intermediate densities (for example, $n_H = 1 \pcc$, shown in the left column) than at lower densities ($n_H = 10^{-4}$, shown in the right column) due to the presence of ionising photons. 
At lower densities, the temperature of the plasma plays a lesser role in determining the ionisation state compared to the ionisation parameter. Therefore for a given radiation field, the ionisation fractions at low densities remain close to the photo-ionisation equilibrium values and are less dependent on the temperature change. The heating is also elevated for lower density but is still low with respect to cooling.
For example, the cooling to heating ratio at $T = 10^5$~K for $n_H = 1$ is $\sim 10^{-5}$, whereas, the same ratio for $n_H = 10^{-4}$ is $\approx 10^{-2}$ (see figure \ref{fig:isochoric-heating-cooling} and also noted in \cite{Oppenheimer2013}).
 The main difference compared to  G17 is our inclusion of statistical charge transfer for ions with charge $\geq 4+$, which mostly affects $\sim 4+, 5+$ ions and, thereafter, propagated to lower ions.  Higher ions ($\gtrsim 5+$) are unaffected as they are not usually present along with \ion{H}{i} or \ion{He}{i} in the plasma. 
 
 The effect of non-equilibrium cooling and heating is shown in figure \ref{fig:isochoric-heating-cooling}. The cooling efficiency (green lines) departs from equilibrium 
 only for temperatures $\lesssim 6\times 10^5$~K. The difference with respect to the 
 photo-ionisation equilibrium case is 
 between a factor of $2$ (near $\sim 10^5$~K)  
 and a factor of $10$ (near $\sim 10^4$~K), i.e. isochoric non-equilibrium cooling is slower than the equilibrium cooling. There is no difference in the cooling efficiencies between G17 and our new results despite the additional charge transfer rate. The heating (shown by red lines), however, is lowered by the introduction of the extra charge transfer at $\sim 10^4$K for $n_H = 1\pcc$. This difference is, however, not visible at lower densities (see right panel) where photo-ionisation dominates.
 
We show the change in ion-fraction ratios \ion{N}{v}/\ion{O}{vi} and \ion{C}{iv}/\ion{O}{vi} with temperature in figure \ref{fig:isochoric-ion-ratio}. The non-equilibrium ion ratios are quite different in comparison to equilibrium ratios below $\lesssim 3\times 10^5$~K. As previously demonstrated, the difference between the G17 and new results only appear at $n_H = 1\pcc$ and $T < 3\times 10^4$~K. In our new computations, the ratios can go to very high values as higher ions like \ion{O}{vi} can now recombine more efficiently through statistical charge transfer. 

\subsection{Radiative transfer}
\label{subsec:tests-rad-trans}
\subsubsection{slanted beam in a sphere}
\label{subsubsec:slanted-beam}
\begin{figure*}
	\centering
	\includegraphics[width=0.7\textheight, clip=true, trim={1cm 3cm 5cm 0cm}]{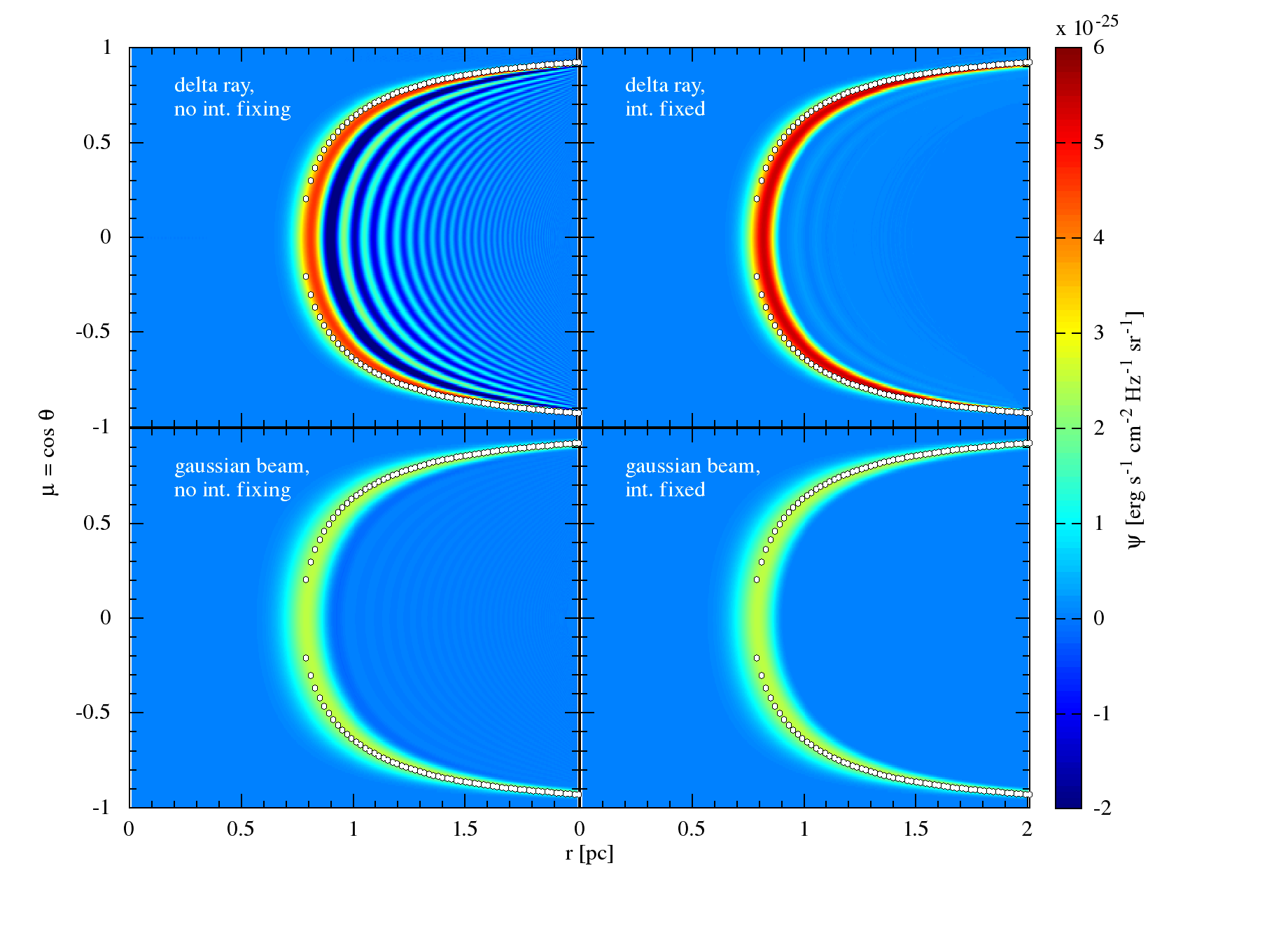}
	\caption{Expected versus the obtained track of a \deltaray (top panel) and a \gaussbeam. The colour shows the sp. intensity. The theoretically expected track is overplotted as the white dots.}
	\label{fig:ray-geometry-sim}
\end{figure*}
\begin{figure*}
	\centering
	\includegraphics[width=0.6\textheight, clip=true, trim={1.2cm 0cm 1.1cm 0cm}]{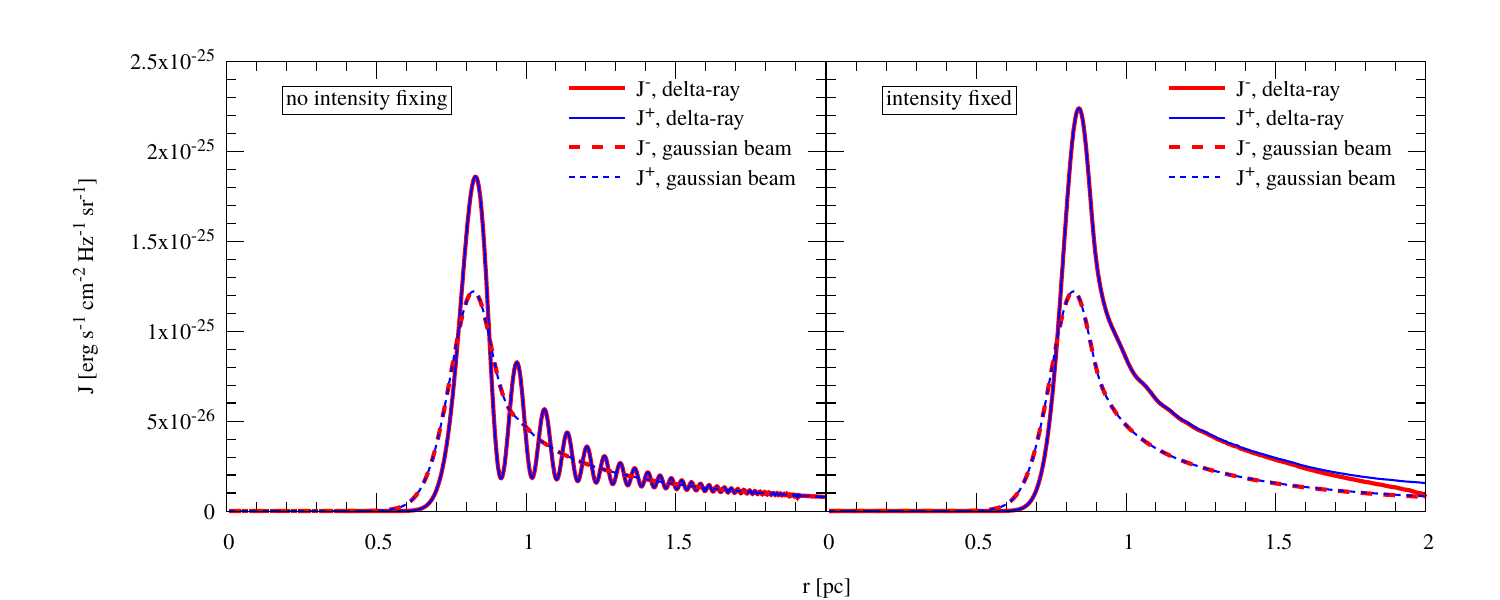}
	\caption{Conservation of angular averaged sp intensity ($J$) along the ray/beam track without neg intensity fixing (left panel) and with negative intensity fixing (right panel). Red represents inwards rays ($\mu<0$), and blue represents outgoing rays ($\mu>0$). In most part of the plot, red is hidden behind the blue lines implying very good energy conservation along the ray/beam.}
	\label{fig:Jnu}
\end{figure*}

To show that our radiative transfer can accurately track the positions of rays inside the simulation box, we inject i) a \deltaray ( $\psi(\mu') \propto \delta(\mu - \mu')$ ) and ii) a \gaussbeam ( $\psi(\mu') \propto \exp\left( - (\mu-\mu')^2/2\omega^2 \right)$ ) at the outer boundary ($r_0 = 2$ pc) of a sphere. Since the outer boundary condition can only be inwards, we choose $\mu' = -0.9219$ and set $\omega$ for the \gaussbeam such that the total energy injected is distributed over only the central $3$ rays.
The ray/beam enters the sphere from outside ($\mu<0$) and passes through the tangent point ($r_{\rm tan} = r_0\sqrt{1-\mu'^2}$) to finally exit via $r_0, -\mu'$. The track of the ray/beam within the sphere is then given as 
\begin{equation}
\mu ( r ) = \sqrt{1 - \left(\frac{r_0}{r}\right)^2\, (1-\mu'^2)} 
\label{eq:delta-ray-track}
\end{equation}
This analytic form of $\mu(r)$ has been compared with the obtained intensity track from the simulations in figure \ref{fig:ray-geometry-sim} where the opacity and emissivity of the sphere are set to be zero. The sphere is discretised in $1024$ and $256$ grid points 
along the $r$ and $\mu$ directions, respectively.

Figure \ref{fig:ray-geometry-sim} shows the $r-\mu$ track of a \deltaray (top panel) and a \gaussbeam (bottom panel). The theoretically expected track (eq \ref{eq:delta-ray-track}) is shown as the white dotted line in each panel. In both 
cases, the track of the ray is well reproduced by the simulations except for a small discrepancy in the \deltaray case. This is because tracking a single ray suffers from limitations due to the linear interpolation method used between the cells.  The mismatch of the theoretically predicted line (Eq \ref{eq:delta-ray-track}) versus the  computed smeared intensity curve (red/yellow region in fig 6) for the \deltaray also disappears once we increase the angular resolution. The discrepancy also disappears as the energy is distributed among a few rays, as can be seen for the \gaussbeam.

Another issue that is immediately apparent in the top left panel of fig \ref{fig:ray-geometry-sim} are the negative intensities and corresponding fringes. As explained earlier (sec \ref{subsec:fix-neg-intensity}), negative intensities arise due to the linear interpolation between the cells and only if the intensity is strongly peaked along one ray. Since the technique to solve RT (sec \ref{subsec:rad-trans}) guarantees energy conservation, a negative intensity in some cells results in excess intensity in nearby rays, which creates the fringes. Notice that these artefacts only appear on the outer side (larger radii) of the predicted ray/beam. This is understandable as the information only propagates from the bottom-right corner to top left corner for $\mu < 0$ and from bottom-left to top-right corner for $\mu >0 $ (see fig \ref{fig:r-mu-grid} and section \ref{subsec:rad-trans}). 

Fortunately, both the negative intensities and the fringes tend to disappear in the \gaussbeam case as soon as the energy is distributed among several rays. We, in any case, employ our negative intensity fixing technique (sec \ref{subsec:fix-neg-intensity})
and the result is shown in the right column of figure \ref{fig:ray-geometry-sim}. For both the \deltaray and \gaussbeam cases, the negative intensity and fringes almost vanish from the map. 
The corresponding energy conservation is shown in figure \ref{fig:Jnu} for both the \deltaray and \gaussbeam cases. The figure compares the angular averaged specific intensity for the incoming ray/beam ($\mu<0$) and the outgoing ray/beam ($\mu>0$). In an ideal case where the energy of the ray/beam is conserved, the averaged intensity for both incoming and outgoing rays should be equal  in the absence of any absorbing/emitting medium. This is exactly what we see in Fig \ref{fig:Jnu}.
Without negative intensity fixing, the angular averaged intensity conserves energy very accurately despite the fringes. With fixing, the energy is conserved very accurately for a \gaussbeam but not for the \deltaray. For the \deltaray, the energy conservation is not very good at the outer radii when the energy is supposed to be only along a single ray. The conservation is much better once the ray travels slightly inwards. 

\begin{figure}
	\centering
	\includegraphics[width=0.35\textheight]{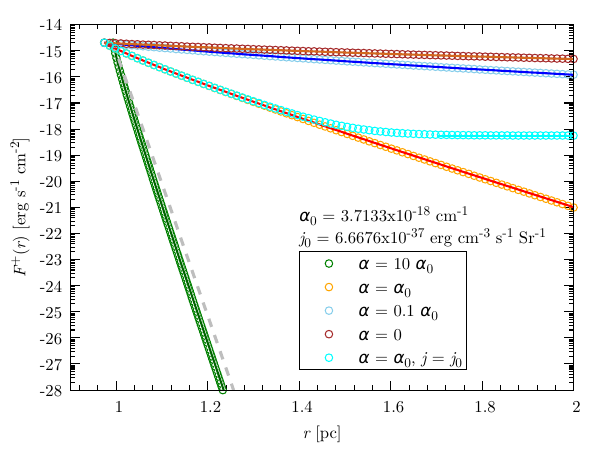}
	\caption{Test for attenuation of flux in a spherical geometry in the presence of  
	opacity. The open circles represent 
	the simulation results, and the solid lines represent the predicted behaviour of the flux (Eq \ref{eq:Fr-predicted}). All the cases have emissivity $j_0 = 0 $, except the cyan points (see text). The dashed gray curve represents a popularly used flux formula, $F(r) \propto r^{-2} \exp\left[- \alpha\:(r-r_c)\right]$, for a radiating surface.}
	\label{fig:attenuation-test}
\end{figure}

\subsubsection{spherical attenuation}
\label{subsubsec:spherical-attenuation}
In this test, we exclude radiation from outside the simulation box and assume that the inner boundary ($r = r_c$) of the spherical grid behaves like a black body surface with brightness $\psi'$.
The outward flux from the central surface is then simply $\pi \psi'$. We set the rest of the simulation box to have no emissivity and a constant absorption coefficient $ \alpha_0 = 3.7133\times 10^{-18} \pcm$ (representing hydrogen at $T \approx 1.5\times 10^4$~K at $n_H = 1 \pcc$ , right after Lyman limit). We also perform tests with varying absorption. The results are shown in Fig. \ref{fig:attenuation-test} and compared with the theoretical curves\footnote{See Appendix \ref{app-sec:flux-from-BB} for the derivation.}.
\begin{equation}
F(r) = 2\: \pi \psi' \int_{\cos(\theta_c)}^{1} e^{-\alpha \left(r\xi - \sqrt{r_c^2-r^2+r^2\xi^2}\right)}\: \xi d\xi 
\label{eq:Fr-predicted}
\end{equation}
where, $\theta_c = \sin^{-1}(r_c/r)$ is the maximum angle at $r$ that contains  the central source (see \cite{ryb+light}, their Fig 1.6). The exponential term appears due to the fact that $\alpha > 0$. For $\alpha = 0$, the above integration would produce the standard result, $F(r) = \pi \psi' \left( r_c/r\right)^2$. However, for a non-zero $\alpha$, one needs to integrate the above equation.  The figure shows a good match with the expected curves 
both at high and low opacities 
over several orders of magnitude. There is, however, a small discrepancy (factor of $\sim 1.5$ over about $10$ orders of magnitude) for $\alpha = 10\: \alpha_0$ which becomes negligible as one considers finer resolution element. 
This brings us to an important property of the RT solver - the size of the radial cells should not be much larger than $1/\alpha$, otherwise the assumption of linear interpolation inside the cells may break and introduce significant error.  

For non-zero emissivity, $j \neq 0$, we do not have a  simple analytical expression for the outgoing flux. However, it is possible to compare the results at large optical depths (measured from the centre) where the specific intensity in any direction approaches the source function ($j/\alpha$) and "forgets" about the central source. We show the test result of such a case (cyan points in figure \ref{fig:attenuation-test}) where we set a constant emissivity ($j = 6.667\times 10^{-37} \ergpspcmsq$ sr$^{-1}$) throughout the medium along with a constant opacity ($\alpha = \alpha_0$). The asymptotic flux, for this case, is then $\pi\times j/\alpha$ which is the solid cyan line. The excellent match with the predicted values shows the compliance of our code to properly account for emissivity as well.

Note that a popular approximation of the flux from a radiating surface in the the presence of an absorbing medium, $F(r) = F(r_c)\:\left(  r/r_c\right)^{-2} \exp\left[-\alpha\:(r-r_c) \right]$ \citep[for example, ][]{Raga2012}, is only valid when the emitted rays from the central surface are radial. One needs to consider the full solution in the form of Eq \ref{eq:Fr-predicted} if the rays are not radial to the surface, as is the case for a stellar surface or a radiating shell. We show the effect of such an approximation in Fig \ref{fig:attenuation-test}. The dashed line shows one such example of the approximated flux from a black-body in a medium with opacity $= 10\:\alpha_0$ and $j = 0$. Although $F(r) \propto 1/r^2 \times \exp\left[-\alpha (r-r_c)\right]$ at larger optical depth, its magnitude is overestimated by a factor of few to an order of magnitude.

\subsection{\strom sphere}
\label{sec:stromgen-sphere}
We now combine the IN and RT schemes to simulate a \strom sphere considering a dynamical evolution of the sphere and interactions with the hydrodynamical variables.
We consider a constant ionising source at the inner boundary ($r= r_c =2.748$ pc) as earlier, but also calculate the \ion{H}{i} abundances and include its opacity. The central ionising spectrum is set to be $\psi_{c,\nu} = 10^7\times J_{\nu, HM12}$ for all the rays with $\mu \geq 0$ while no radiation is set to enter the sphere from its outer boundary\footnote{Note that the use of the HM12 spectrum at the inner boundary is completely arbitrary. This is just to test the effectiveness of the code and does not implicate any realistic physical scenario.}. Therefore, the flux at $r = r_c$ is given by $\pi\: \psi_{c, \nu}$. 
Given the HM12 spectrum, this produces 
an HI ionising flux ($> 13.59$ eV) of $Q = 2.16\times 10^{49}$ s$^{-1}$. 
We set the density of the sphere to $n_H = 1 \pcc$ (with no He or metals) and the temperature to $6\times 10^3$~K. We also switch off any direct radiation pressure on the atoms but do include photo-heating. 

\begin{figure*}
	\centering
	\includegraphics[width=\textwidth]{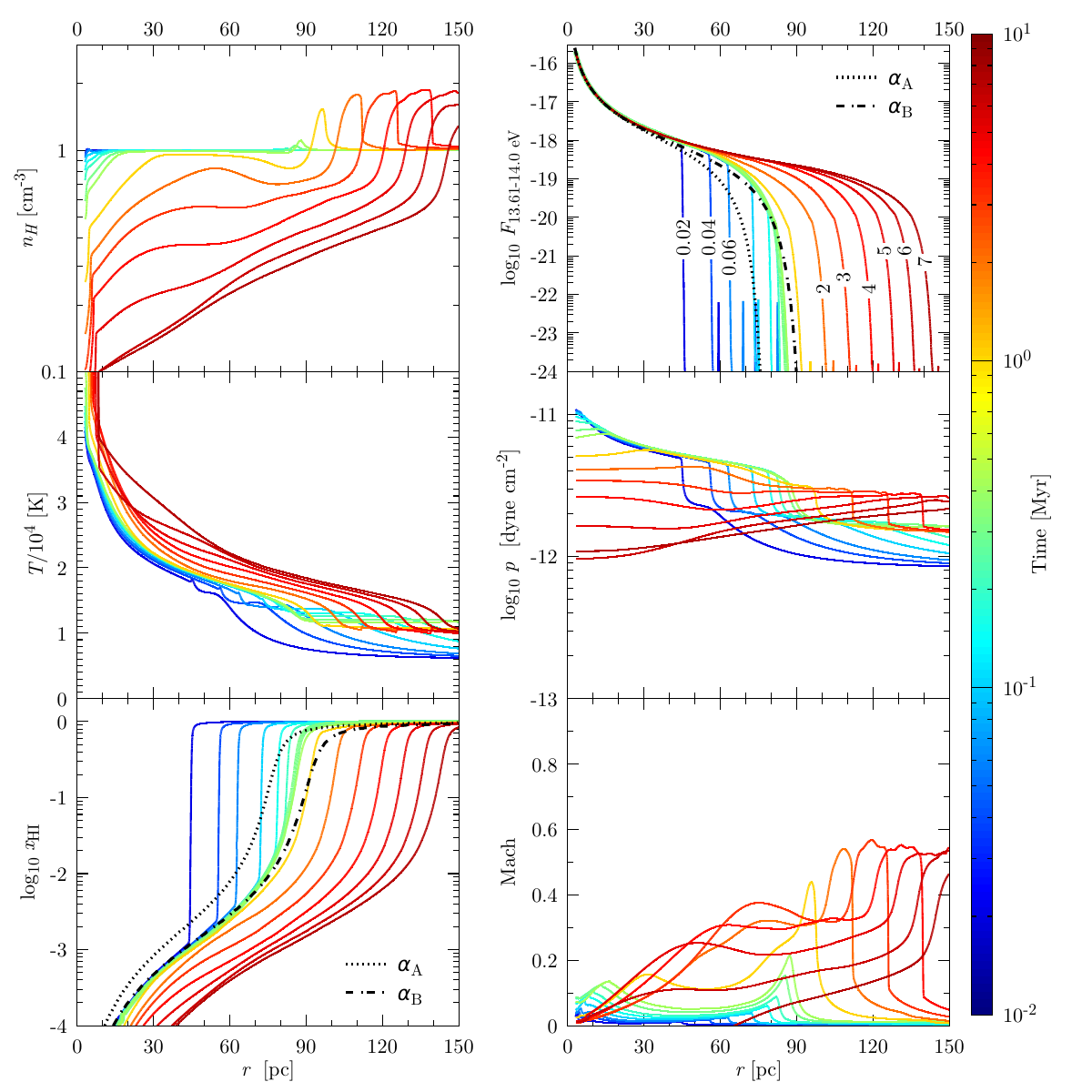}
	\caption{Structural evolution of a \strom sphere with time (represented by different colours). Quantities shown are, Hydrogen number density $n_H$ (top-left panel), temperature (centre-left panel), Hydrogen ionisation fraction, $X_{\rm HI}$ (lower-left panel), the radial flux ($\ergpspcmsq$ Hz$^{-1}$ ) in a $13.61-14$ eV band (top-right panel), thermal pressure (centre-right panel) and the Mach number (lower-right panel). The dotted and dash-dotted black lines represent the theoretical profiles assuming static, uniform density `case A' or `case B' recombination, respectively. These profiles should be compared with the thick green lines (in $x_{\rm HI}$ and Flux panels) which represent the profiles at $t = 300$ kyr $\simeq 2\times t_{\rm rec, H II}$, where $t_{\rm rec, H II}$ is the set-up timescale for the\strom sphere at $T = 1.5\times 10^4$ K. The case-B, "on-the-spot" approximation over-estimates the \strom sphere radius relative to the partially-thick numerical solution.}
	\label{fig:strom-struct}
\end{figure*}

The time evolution of the ionisation and dynamical structures are shown in figure~\ref{fig:strom-struct}. The evolution depicted in this figure can be divided into two stages: an early stage and a late stage. The early stage is driven by the evolution of the ionisation front and ends by forming the classical \strom sphere, after about a recombination time, and a late-stage, where pressure gradients (from photo-heating) drive the dynamical evolution. This part is shown by the ``cold"-colour-curves in Figure~\ref{fig:strom-struct}, and is the focus of section \ref{subsubsec:evol-of-ion-front}. The late-stage is driven by the pressure gradients (from photo-heating) drive dynamical evolution, which further increases the size of the ionised region. This part is shown by ``warm"-colour-curves and is discussed in section \ref{subsubsec:strom-expansion}.

\subsubsection{Evolution of ionisation front}
\label{subsubsec:evol-of-ion-front}
Initially, the ionisation front (IF) propagates through the medium, until after a recombination time, the radial structure of the ionisation fraction and the radiation flux  
reach a steady state over a time-scale close to the recombination time-scale of \ion{H}{ii} ($t_{\rm rec, H II}$). It is apparent that the \ion{H}{i} fraction roughly anti-correlates with the flux. 

The ionisation front is terminated at a radius roughly close to the expected \strom sphere $R_{\rm st} \approx 83 $ pc. We also notice that the IF is not sharp but falls off slowly at larger radii. This shallow tail also helps to pre-heat the material ahead of the IF, as can be seen in figure \ref{fig:strom-struct}.

The centre-left panel of Figure \ref{fig:strom-struct} shows the temperature distribution at different time snapshots. It shows that the gas is heated to a few times $10^4$~K inside the IF due to photo-heating. The heating is not only restricted to the inner region of the IF but also extends beyond the IF. This heating increases the temperature of the background medium from $6\times 10^3$~K to $\approx 10^4$~K before the passage of the IF through it.

\begin{figure}
	\centering
	\includegraphics[width=0.45\textwidth, clip=true, trim={0cm 0cm 1cm 1cm}]{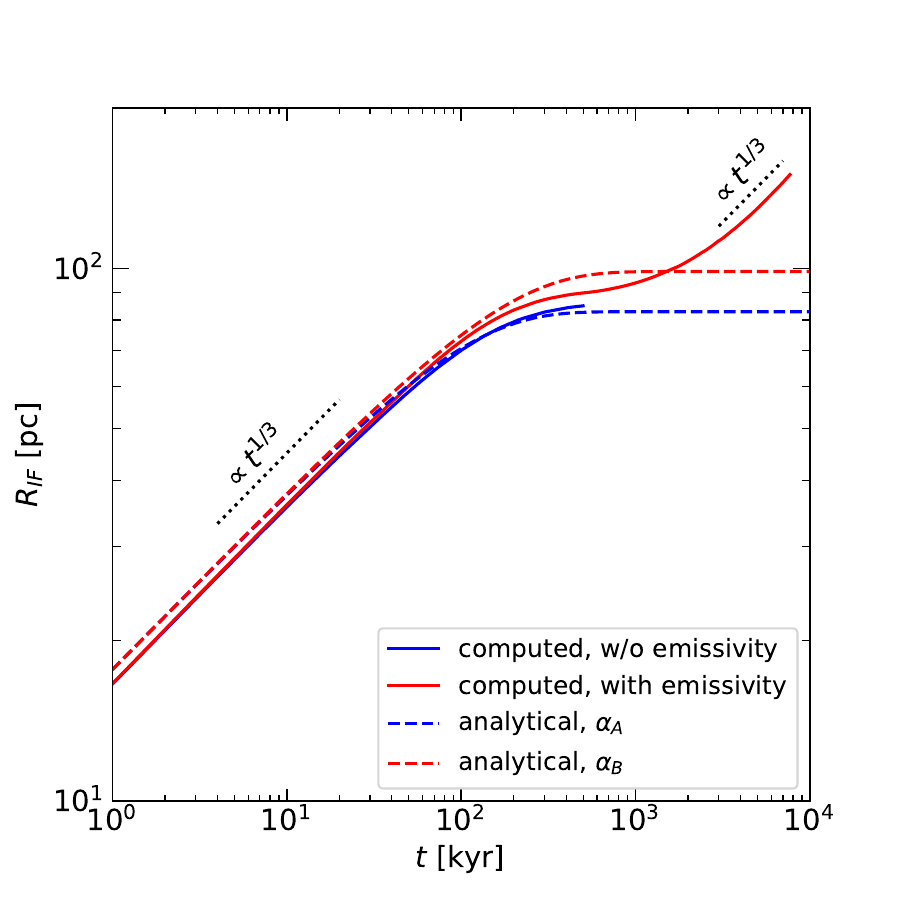}
	\caption{Evolution of the ionisation front (solid lines) for a ionising luminosity of $Q = 2.16\times 10^{49}$ s$^{-1}$ and hydrogen density of  $n_H = 1.0$.  
	Blue represents the simulation with zero emissivity, whereas red shows the case with emissivity turned on. The dashed lines show the corresponding theoretical profiles for both  cases. The rising red line at $t \sim 1$ Myr shows the start of the dynamical expansion of the over-pressurised bubble. The dotted black lines show reference power laws.}
	\label{fig:ionis-front}
\end{figure}

We perform two kinds of \strom sphere tests.
In the first case, we switch off the local emission at every cell (but include radiation losses), which means that any photon that is emitted by a recombining plasma is not re-absorbed. This is similar to case A recombination in an optically thin medium with coefficient $\alpha_A$. In the second case, we turn on the local emission from the plasma and allow the emitted photons to be re-absorbed, depending on the optical depth. This allows the gas to approach "case B" recombination.
The expected sizes of the \strom spheres for pure case-A and B are
\begin{eqnarray}
R_{\rm st, A} &=& \left( \frac{3\:Q}{4\pi n_H^2 \alpha_A}\right)^{1/3} \approx 83 \mbox{  pc} \nonumber \\
R_{\rm st, B} &=& \left( \frac{3\:Q}{4\pi n_H^2 \alpha_B}\right)^{1/3} \approx 98 \mbox{  pc}
\label{eq:Rst_AB}
\end{eqnarray}
Where we have assumed uniform density and a fixed temperature of the recombining plasma, of 
$1.5 \times 10^4$~K following figure \ref{fig:strom-struct} for both the cases and $\alpha_A = 4.13\times 10^{-13} T_4^{-0.7131}$ cm$^3$ s$^{-1}$ and $\alpha_B = 2.56\times 10^{-13} T_4^{-0.8163}$ cm$^3$ s$^{-1}$ \citep{Draine2011}.

The theoretical profiles 
for any of the assumed recombination rate, can be obtained by solving  \citep[Eq 2.18 of][]{Osterbrock2006}
\begin{equation}
\frac{(1-x_{HI})^2}{x_{HI}} = \frac{1}{n_H \alpha(T)}\: \frac{r_c^2}{r^2}\: \int_{\nu_0}^{\infty} \frac{F_\nu (r_c)}{h\nu}\: \exp^{-\tau_\nu (r)} a_\nu\: d\nu
\end{equation}
where $\nu_0 = 3.29\times 10^15$ Hz represents the Hydrogen ionisation potential, $n_H = n_{H I} + n_{H II}$ is the total hydrogen density inside the sphere, $\tau_\nu(r) = n_H \sigma_{\nu, H}^{\rm pi} \int_{r_c}^r x_{H I}(r') dr'$ is the optical depth from the centre to a given radius $r$ and $\sigma_{\nu, H}^{\rm pi}$ is the shell integrated photo-ionisation cross-section for \ion{H}{i}. The radial flux can be calculated using $F_{\nu}(r) = F_\nu(r_c)\: (r_c/r)^2\: \exp(-\tau_\nu(r))$. Note that this form of the radial flux is valid only if either the opacity is negligible or 
the rays are purely radial. For a general optically thick medium with uniform opacity see eq \ref{eq:Fr-predicted}. Fortunately, for the \strom sphere, this form of the flux is valid due to the negligible opacity  in the inner region and almost radial rays in the outer parts where the opacity is not negligible.

Figure \ref{fig:strom-struct} shows the theoretical profiles for the ionisation and radial flux for $T = 1.5\times 10^4$ K and $n_H = 1$. The dotted and dash-dotted lines show the theoretical profiles for 
pure, uniform density `case A' and 'case B' recombination, respectively. We find that the profile for the assumed `case B' recombination is quite consistent with the results obtained from our code.  The theory, however, slightly over-predicts the radius of the \strom sphere.

This discrepancy is also shown in figure \ref{fig:ionis-front} in a more quantitative way. The figure compares between the theoretical and numerically obtained ionisation front radii (taken as the radius where $X_{\rm H I} = 0.5$). 
The expected expansion of the ionisation front ($R_{\rm IF}$) is given by 
\begin{equation}
    R_{\rm IF}^3 = R_{\rm st}^3\,\exp\left[ - n_H \alpha(T)\: t \right]
\end{equation}
where, $R_{\rm st}$ is the \strom radius for a given recombination coefficient $\alpha(T)$ as noted in Eq \ref{eq:Rst_AB}. 
The figure shows a good match between the "no-emissivity" model (blue solid) and the analytical model for "case A" (blue-dashed) before the over-pressurised bubble dynamics becomes important.
When emissivity is included, the size of the ionised region is slightly smaller than expected for "case B" (red).
We speculate that this underestimation of $R_{\rm IF}$ in the $\alpha_B$ case may be for the several reasons. Most importantly, the assumption of the case B recombination rate while performing the analytical estimation. This recombination rate is accurate only if the optical depth is $\gg 1$ (`on-the-spot' absorption) unlike the \strom sphere where the optical depth is $\sim 1$. The actual recombination rate for a \strom sphere, therefore, should be between $\alpha_A$ and $\alpha_B$. Other reasons include the use of band  
averaged opacities which are weighed by the UV background. This may underestimate or overestimate the actual instantaneous opacity depending on the hardness of the local spectrum compared to the UV background. It is also possible that the average temperature inside the ionised bubble for the case when we turn on emission from gas is slightly lower than compared to the zero emissivity case. 
Additionally, the analytical estimation for the \strom sphere depends on the assumption of a constant density, temperature and a sharp boundary for the ionised sphere. In reality, none of these assumptions are true as can be seen in Fig \ref{fig:strom-struct}. 
In addition, we also estimated \strom radius by equating the total recombination rates in these two cases with the $Q$ from the central star. These radii are about $106$ pc and $118$ pc for the case of zero emissivity and full emissivity, respectively. These radii, once used in \strom radii calculation (Eq \ref{eq:Rst_AB}) also indicate smaller average temperature inside the ionised sphere of the full emissivity case. Hence, we do not consider the underestimation of $R_{\rm st,B}$ as a drawback to the simulation, rather a success of the test.
\subsubsection{Expansion of the \ion{H}{ii} region}
\label{subsubsec:strom-expansion}
Our discussion of the \strom sphere so far does not account for the thermal pressure of the ionised sphere. In reality, the ionised gas is heated by photo-heating and the pressure further increases by   
the increase in the mean particle number of the ionised plasma. It can be easily estimated that the thermal pressure of the \strom sphere, in our case, will be only about a factor of a few higher compared to the background (a factor of $2$ due to the increasing particle number and a factor of $\sim 2$ due to photo-heating). This can also be seen in the centre-right panel of fig \ref{fig:strom-struct}. Clearly, the initial \strom sphere is over-pressurised, and gas dynamics should be accounted for.

The \strom sphere takes about $t \sim 2\:t_{\rm rec, H II} \approx 300$ kyr to set-up a static ionised sphere, whereas, the sound crossing time is $\sim R_{\rm st,B}/\sqrt{5/3 k_B T/\mu m_p} \approx 4.7$ Myr. Therefore, the gas dynamics for the ionised sphere can be safely neglected for $t\lesssim $ few$\times t_{\rm rec, H II}$. However, the dynamics becomes important after this time as can be seen in Fig \ref{fig:strom-struct} and \ref{fig:ionis-front}. Figure \ref{fig:strom-struct} shows the late-time evolution of the ionised sphere (in the reddish shade). The most notable effect of the over-pressurised bubble is that it sweeps up the matter inside the ionised sphere to a somewhat thinner shell, propagates outwards, and finally exits the computational box. The density interior to the shell decreases almost by a factor of few, thus practically creating a bubble (see the top-left panel). The dynamics of such over-pressurised bubbles have been studied in the literature \citep{Spitzer1968, Dyson1980, Raga2012}.

Fig \ref{fig:ionis-front} also shows the late-time evolution of the ionisation front (which is almost coincidental with the bubble radius). We notice that the expansion of the ionised sphere really picks up only after $t \gtrsim 1$ Myr. Although the theoretical expansion of the bubble is $R_{\rm IF} \propto t^{4/7}$ \citep{Krumholz2009}, the simulated bubble only reaches $R_{\rm IF} \propto t^{1/3}$ in this regime before it exits the simulation box.

%
%
%
\section{Discussion}
\subsection{Emission spectra of \strom sphere}
We also present a tool to calculate the emergent spectra from the sphere. Our tool contains a separate script to solve the frequency-dependent radiative transport at a given time. The  RT method is the same as presented in section \ref{subsec:rad-trans} but with a much higher frequency resolution suitable to 
include the impact of line emission. The emissivities and opacities required to perform the RT are obtained from \cloudy-17 by using the local density, temperature and non-equilibrium ion-fractions. The spectra, therefore, may contain signatures of non-equilibrium ionisation for comparison to observations and predictions.
 \begin{figure}
 	\centering
 	\includegraphics[width=0.5\textwidth, clip=true, trim={0cm 0cm 0cm 0.5cm}]{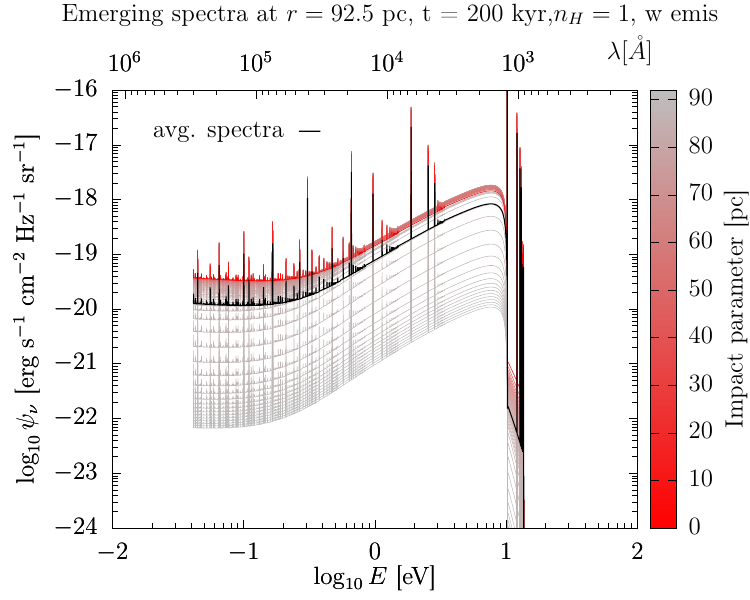}
 	\caption{Synthetic spectra from a \strom sphere at $t = 200$~Kyr emerging from a radius of $r = 92.5$ pc for $n_H = 1$, including emission but without including any dust. The colour of each line represents the impact parameter from the centre of the SN remnants. The deep black spectrum is the surface averaged value of all the spectra.}
 	\label{fig:spectra}
 \end{figure}
The emergent spectra, at $r = 92.5$ pc, where $x_{HI} = 0.95$ at $t = 200$~kyr, is shown in Fig \ref{fig:spectra} as a function of impact parameter from the centre of the ionised sphere. The impact parameters ($b$) plotted in this figure are simply converted from the $\mu$ values at that radius since $b = r \sin\theta = r \sqrt{1-\mu^2}$. The deep black line shows the surface averaged spectra which is same as the angle averaged spectra, $<\psi> = \int_0^1 \psi(\mu) d\mu = \sum_{m = N_\mu/2}^{m = N_\mu} \frac{\Delta \mu}{2} \left( \psi_m + \psi_{m+1} \right)$, in case the remnant is not resolved. The sudden rise in emission and drop thereafter at $\lambda = 1216\, A^\circ$ is due to the Ly-$\alpha$ emission and scattering (since the scattering is treated as absorption in the first step but is considered as emission in the next time step). The final drop of emissivity happens at $\lambda\leq 912\, A^\circ$ ($E \geq 13.59$ eV) due to the neutral H absorption.

\subsection{Effect of dust}
\begin{figure}
	\centering
	\includegraphics[width=0.4\textwidth, clip=true, trim={0cm 0.5cm 1.3cm 1.5cm}]{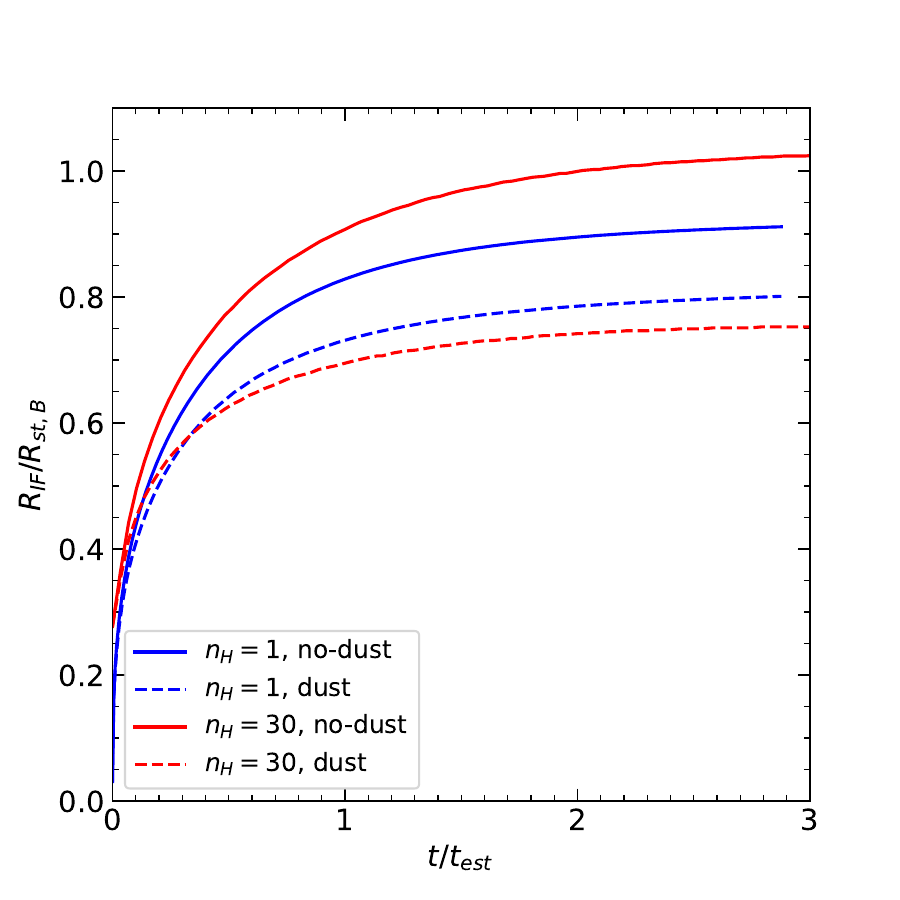}
	\caption{Evolution of ionisation front for a \strom sphere test with (dashed lines) and without (solid lines) dust. The result has been shown for two densities, $n_H = 1 \pcc$ (blue) and $n_H = 30 \pcc$ (red).}
	\label{fig:strom-dust}
\end{figure}

To examine the effect of dust absorption we run the simulations in section \ref{sec:stromgen-sphere} with and 
without dust at two different densities, $n_H = 1$ and $30$ cm$^{-3}$. The results are  
shown in figure \ref{fig:strom-dust}. The figure shows the evolution of the ionisation fronts normalised by their corresponding theoretical \strom radius, $R_{\rm st, B}$ and its recombination time, $t_{\rm rec} = 1/n_H \alpha_B(T)$ at $T = 1.5\times 10^4$~K.
The figure demonstrates the effect of dust in a denser medium.  We find $\approx 30\%$ decrease in the final radius for $n_H = 30 \pcc$ compared to only $\approx 13\%$ decrease for $n_H = 1 \pcc$. This verifies our discussion regarding the effect of dust in denser medium (Eq \ref{eq:nH-dcrit}) and shows that our code is well suited for the studies where dust plays a major role.
\begin{figure*}
	\centering
	\includegraphics[width=\textwidth, clip=true, trim={0cm 0cm 1.3cm 0cm}]{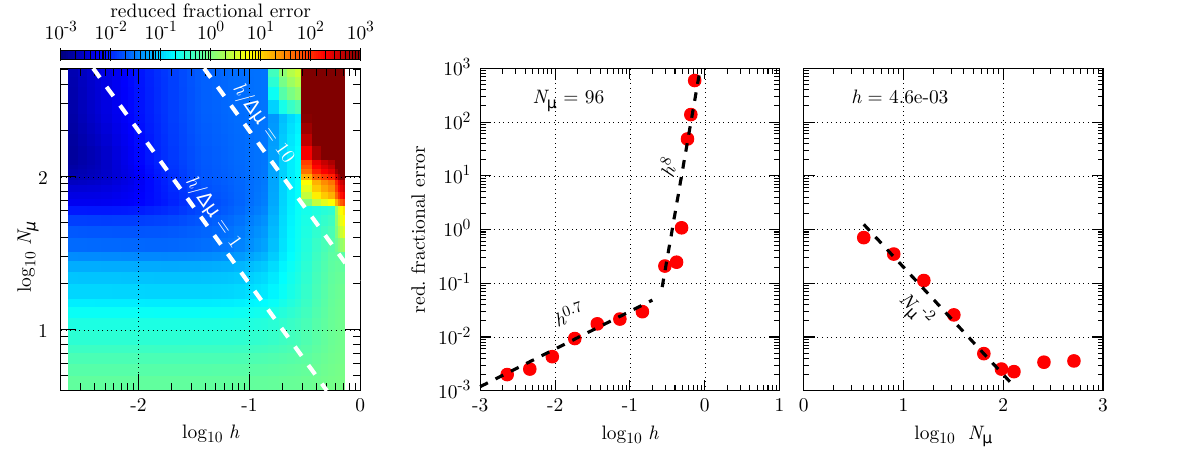}
	\caption{Reduced fractional error in our RT scheme for the spherical attenuation case as a function of typical optical depth per cell, i.e., $h = \alpha \Delta r/2|\mu|$. We assume $\mu = 1$ since most of this error comes from the $\alpha \Delta r \gg 1$ where the rays contributing to the flux is already close to $\mu \sim +1$. The dashed white lines in the left panel represent the possible error caused by the rays that lie close to the tangential plane ($|\mu|_{\mbox{min}} = \Delta \mu = 2/N_\mu$) at any radius and are determined by the number of rays. The middle and right panels show the same data as in the left panel except to show the dependencies on either $h$ or $N_\mu$. The dashed black lines in the middle and right panels show reference power laws to guide the eyes.}
	\label{fig:rt-error}
\end{figure*}

\subsection{Error in the RT scheme}
Numerical error is unavoidable in any numerical schemes. Our RT scheme is no exception. The convergence of the scheme is shown in figure \ref{fig:rt-error} for the case of the spherical attenuation test discussed in section \ref{subsubsec:spherical-attenuation}. The figure shows the reduced fractional error
\begin{equation}
\epsilon = \frac{1}{N_r} \sum_{i=1}^{N_r} \frac{|F_{\rm theo}(r_i)-F_{\rm num}(r_i)|}{F_{\rm theo}(r_i)}\,\,,
\label{eq:error}
\end{equation}
where $F_{\rm theo}(r_i)$ is the theoretical flux obtained from Eq \ref{eq:Fr-predicted} and $F_{\rm num}(r_i)$ is the value obtained from our RT scheme. This definition of the error makes sure that the absolute value as well as the general shape of the curve are  
taken into account. The theoretical and numerical fluxes are compared only to $r = 3$ pc where the optical depth is $\approx 23$ (assuming $\alpha = 3.8\times 10^{-18}$ cm$^{-1}$) and the absolute flux decreases by $\sim 10$ orders of magnitude. The presented error therefore is a very strict one and a somewhat higher value, using this definition, does not necessarily mean the scheme is unusable for practical purposes. 

The left panel of figure \ref{fig:rt-error} shows\footnote{The  error check in practice is done by increasing $N_r$ or $N_\mu$ by a factor of 2. The intermediate points are obtained by a linear surface interpolation of the actual data to make it smooth.} $\epsilon$ as a function of $h \equiv \alpha \Delta r$/2 and number of rays $N_{\mu}$. Therefore, $h$ in this plot represents the optical depth along the radial direction. Its definition 
follows from section \ref{sec:num-rt-technique} with the assumption of $|\mu_m| = 1$ considering that most of the error comes from the high optical depth regions where the rays are almost radial. The figure shows that $\epsilon$ decreases as we either decrease $h$ or increase $N_\mu$. We described earlier, the RT scheme best works when $h \lesssim 1$. This scheme also breaks down for very high values of $N_\mu$ when $h \sim 1$. This is because high values of $N_\mu$ samples rays close to $\mu \sim 0$ for which the optical depth even for a single grid cell is very large. This leads to a high value of error. The number of rays above which the error becomes large if given by $h/\Delta \mu \equiv h N_\mu/2 \gtrsim 1$. This limit is shown by one of the white dashed lines in the left panel. However, since the calculation of radial flux from the specific intensity along the rays is weighted by $\mu$, this constraint is a bit less stringent, and  practically appears only at $h/\Delta \mu \gtrsim 10$.

The convergence of $\epsilon$ with $h$ is shown in the middle panel of the figure \ref{fig:rt-error}. It shows that although our RT scheme converges very quickly for $h\gtrsim 1$, the convergence is only sub-linear ($h^{0.7}$) in the more sensible regime, i.e. $h \lesssim 1$. The slower convergence of $\epsilon$ compared to the expected 2nd-order convergence is presumably 
due to the high error introduced by the rays close to $\mu=0$ in the inner radii. The error, however, converges quite fast ($\sim N_\mu^{-2}$) with the number of rays, as is shown in the right-hand side panel. However, for a very high number of rays (for a given $h$), the error does not improve due to the extra error carried by the tangential rays ($\mu \sim 0$). Despite a slow convergence with $h$ and a floor in $N_\mu$, the error in our RT scheme is very small for reasonable values of $N_r$ and $N_\mu$. We emphasise that most of the error comes from the high optical depth regions where the flux is already several orders of magnitude smaller than starting value and, therefore, a reduced error of even $\epsilon \sim 1$ is also practically acceptable for realistic calculations.

\subsection{Computational cost}
\label{subsec:compute-time}
\begin{figure*}
    \centering
    \includegraphics[width=\textwidth, clip=true, trim={2cm 0cm 3cm 0cm}]{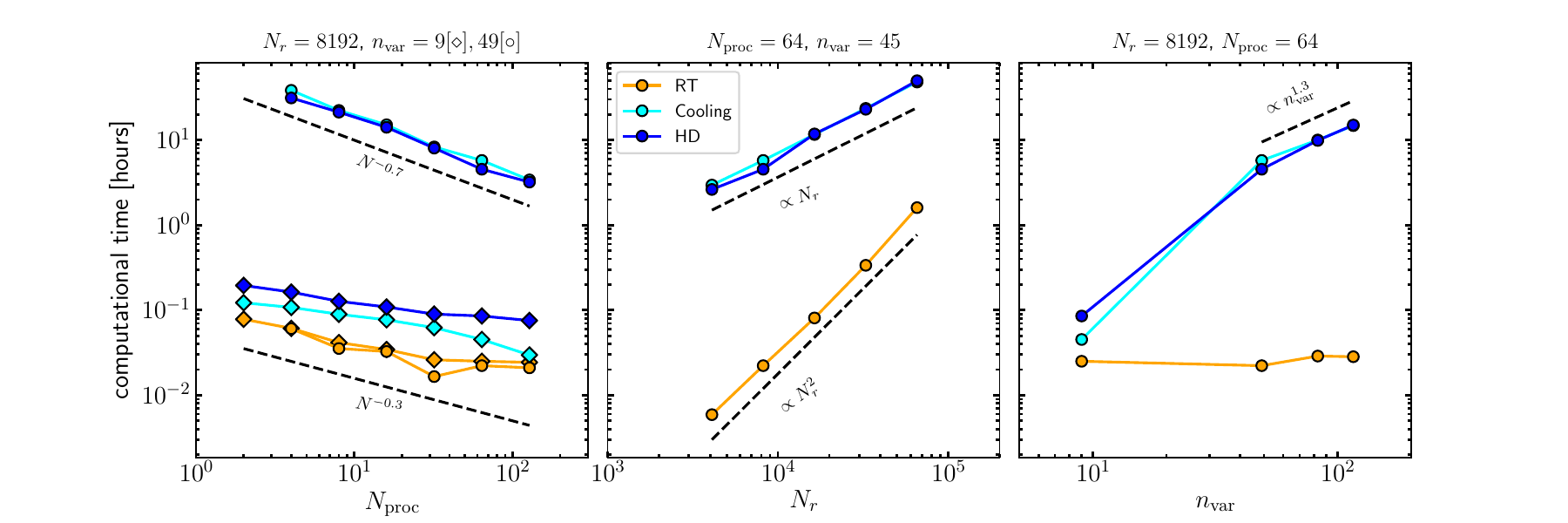}
    \caption{Computational time for solving the combined code to evolve till $100$ kyr for the \strom sphere. Different components of the code are shown using different coloured line-points. The left panel represents the computational time for two categories of simulations, one with only H and He ($n_{\rm var} = 9$; diamonds), and another with all the metals with maximum ionisation level considered is $4+$ ($n_{\rm var} = 49$; circles). The dashed black lines represent typical power laws to guide the eyes.}
    \label{fig:runtime}
\end{figure*}

The computational cost for the different components of the code is shown in figure \ref{fig:runtime}. The time represents the computation time for the code to evolve the \strom sphere (described in section \ref{sec:stromgen-sphere}) till $100$ kyr. A fixed  
physical time is chosen instead of a fixed number of steps 
to account for the varying step-size in \pluto depending on the local Courant–Friedrichs–Lewy condition. The left panel of the figure shows speed up against parallel processing ($N_{\rm proc}$ is the number of processors) for two categories of simulations, one where the NEI network only contains H and He, i.e. the total number of variables to solve for is $n_{\rm var} = 9$ (density, velocity, pressure, 2 tracers, \ion{H}{i}, \ion{He}{i}, \ion{He}{ii} and \ion{He}{iii}), and the second, where the metals are also considered but the maximum level of ionisation is assumed to be $4+$ (thus $n_{\rm var} = 49$). We see that the parallelisation in the case of $n_{\rm var} = 9$ is not great. The $n_{\rm var} = 49$ case can, however, be reasonably parallelised (left panel). In both 
cases, the time to solve the RT algorithm is significantly shorter than the time to solve hydrodynamic or cooling equations. 
It is also seen that the HD step and the cooling steps (includes calculation of the NEI network) take almost equal time since both these parts need to solve either tracer equations (during HD steps) or explicit-time integration (for solving NEI network).

The middle panel shows the computation time with an increasing number of radial grid cells, $N_r$ for 
fixed $N_{\rm proc}$ and $n_{\rm var}$. The HD 
and cooling times follow a simple $\propto N_r$ rule. However, since the time steps in \pluto decrease almost linearly with decreasing cell size (due to CFL constraint), the total number of steps to complete $100$ kyr increases as $\propto N_r$. Now, given that the RT algorithm is performed once every step (unlike the HD or cooling time which have adaptive step sizes) and that the number of grid cells also increases, the time to perform RT increases as $\propto N_r^2$. In the right-hand side panel, we  
show the time taken by the simulation if we trim the NEI network at different levels\footnote{We do not have any points for $n_{\rm var} \sim 20$ since it will require the network to consider only till $+1$ ionisation level, where the network trimming does not work. See sec \ref{subsec:cooling-heating} for more details.}. As expected, the computation time increases almost linearly with the number of variables considered. A significantly less time for the `H and He only case' is achieved by cutting the metal networks completely and thereby skipping some unnecessary computation.

We also note that since the RT scheme takes significantly 
less time than the HD and cooling components of the code, a general multidimensional RT method  is not expected to be the main bottleneck in terms of the computational expense.

\section{Conclusions}
We have presented a new module for \pluto-4.0 that contains an upgraded ionisation network for almost all the important metals, their ionisation states and their contribution to cooling. This network is also coupled to a frequency-dependent radiative transfer module that calculates the local intensities and flux on-the-fly, assuming spherical symmetry. We also employ some dust physics to account for dust attenuation of the ionising radiation field.

We present several tests to demonstrate the accuracy of the ionisation network and radiative transfer, both individually and when coupled together. This is a major upgrade from the previously existing but smaller ionisation network. Although the radiation transfer module works only in spherical symmetry compared to the recent development of 3D radiative transfer module in \pluto, our RT module is frequency-dependent and employs a discrete ordinate technique (short-characteristic) that does not use an \textit{ad-hoc} Eddington tensor to close the moment equations.
Moreover, the employment of the short characteristic method enables us to accurately compute the densities of ions that occur near the boundary between optically thin a thick medium. In addition, the multi-frequency approach enables the calculation of very accurate ionisation rates of different elements and their ions without assuming a single opacity/emissivity for all of them. 

This module is suitable for studying systems with no radiation as well as the ones with highly varying radiation fields in both time and space. One such example is shown in section \ref{sec:stromgen-sphere} as a standard test. In a companion paper \citep{Sarkar2020b} we use our new tool to study the time evolution of heavy element column densities in (non-steady state) expanding supernova bubbles and shells.
 We hope that this tool will help us understand many unanswered questions in astrophysics and will prove to be a powerful tool to the community to better predict the resulting metal column densities and emission spectra of a numerical simulation. 

\section*{Acknowledgements}
We thank Yuval Birnboim, Chi-Ho Chan, Yakov Faerman and Eli Livne for helpful discussions while preparing the code. We also thank the anonymous referee whose comments help us improve the quality of this paper. This work was supported by the Israeli Centers of Excellence (I-CORE) program (center no. 1829/12), the Israeli Science Foundation (ISF grant no. 2190/20),
and by the German Science Foundation via DFG/DIP grant STE 1869/2-1 GE625/17-1 at Tel-Aviv University. We thank the Center for Computational Astrophysics (CCA) at the Flatiron Institute Simons Foundation for hospitality and computational support via the Scientific Computing Core.

\section*{Data availability}
The data underlying this article are available in GitLab at \url{https://gitlab.com/kartickchsarkar/pluto-neq-radiation}

\bibliographystyle{mnras}
\bibliography{library}


\appendix
\section{Radiative transfer method}
\label{app-sec:rad-trans-method}
Starting from eq \ref{eq:RTE-sph}, we can remove the differentials by integrating it first in $\mu$-direction and then in $r$ direction within a $r-\mu$ cell. Integrating it from $\mu$ to $\mu_{m+1}$ we obtain 
\begin{eqnarray}
&&\frac{1}{r^2}\, \frac{\partial}{\partial r}\left( r^2\, \int_{\mu_m}^{\mu_{m+1}}\mu\: \psi\: d\mu \right) + \frac{1}{r} \left[ (1 - \mu^2)\: \psi \right]_{\mu_m}^{\mu_{m+1}} \nonumber \\
&=& j\int_{\mu_m}^{\mu_{m+1}} d\mu  - \alpha\, \int_{\mu_m}^{\mu_{m+1}} \mu\: \psi\: d\mu
\end{eqnarray} 
At this stage we assume that $\psi$ is linear between any $\mu_m$ and $\mu_{m+1}$, i.e.
\begin{eqnarray}
\psi(\mu) &=& \frac{\psi_{m+1}-\psi_{m}}{\mu_{m+1}-\mu_m}\: (\mu - \mu_m) + \mu_m \nonumber \\
&=& \frac{\psi_{m} \mu_{m+1}-\psi_{m+1} \mu_m}{\mu_{m+1}-\mu_m} + \frac{\psi_{m+1}-\psi_{m}}{\mu_{m+1}-\mu_m}\: \mu 
\label{app-eq:psi-interp}
\end{eqnarray} 
Now we can write down different moments of $\psi$ as  
\begin{eqnarray}
\int_{\mu_m}^{\mu_{m+1}}\psi\:d\mu &=&  \frac{\Delta \mu}{2}\: \left(\psi_{m}+\psi_{m+1}\right) \nonumber \\
\int_{\mu_m}^{\mu_{m+1}} \mu\:\psi\:d\mu &=& A_m\:\psi_m - \bar{A}_m\:\psi_{m+1} \nonumber
\end{eqnarray}
where,
\begin{eqnarray}
\Delta\mu_m &=& \mu_{m+1}-\mu_m  \nonumber \\
A_m &=& \frac{\Delta\mu}{6}\:(\mu_{m+1}+2\mu_m) \nonumber \\
\bar{A}_m &=&  - \frac{\Delta\mu}{6}\:(2 \mu_{m+1}+\mu_m)
\label{app-eq:Am}
\end{eqnarray} 
The RTE then becomes 
\begin{eqnarray}
\hspace{-4mm}\frac{1}{r^2}\: \frac{\partial}{\partial r}\left[r^2\left( A_m\psi_m - \bar{A}_m\psi_{m+1}\right)\right]\hspace{-3mm} &+& \hspace{-3mm} \frac{1}{r} \left[ (1 - \mu_{m+1}^2)\: \psi_{m+1} - (1 - \mu_{m}^2)\: \psi_{m} \right] \nonumber \\
 &=&\hspace{-3mm} -\alpha\, \frac{\Delta\mu}{2}\: \left(\psi_{m+1}+\psi_m \right) + j\Delta\mu \nonumber \\
\Rightarrow \frac{\partial}{\partial r}\left[r^2\left( A_m\psi_m - \bar{A}_m\psi_{m+1}\right)\right] \hspace{-3mm} &+& \hspace{-3mm} \left[\left(1-\mu_{m+1}^2\right)\: r + \frac{\alpha\: \Delta\mu}{2}\:r^2 \right]\:\psi_{m+1} \nonumber \\
&-&\hspace{-3mm} \left[\left(1-\mu_{m}^2\right)\:r - \frac{\alpha\: \Delta\mu}{2}\:r^2 \right]\:\psi_m \nonumber \\
&=& j\:\Delta\mu \: r^2 
\end{eqnarray}
We can now continue to use this method and integrate the above equation between two radial grids $r_i$ and $r_{i+1}$ assuming that $\psi$ can be linearly interpolated between the radial grids too. The moments are
\begin{eqnarray}
\int_{r_i}^{r_{i+1}} r^2 \psi\: dr &=& B_i\: \psi_i + \bar{B}_i\:\psi_{i+1} \nonumber \\
\int_{r_i}^{r_{i+1}} r \psi\: dr &=& C_i\: \psi_i + \bar{C}_i\:\psi_{i+1}
\end{eqnarray}
where
\begin{eqnarray}
B_i &=& \frac{1}{12\: \Delta r}\: \left( r_{i+1}^4 - 4\:r_i^3\:r_{i+1} + 3\:r_i^4\right) \nonumber \\
\bar{B}_i &=& \frac{1}{12\: \Delta r}\: \left(3 r_{i+1}^4 - 4\:r_i\:r_{i+1}^3 + \:r_i^4\right) \nonumber \\
C_i &=& \frac{\Delta r}{6} \left(r_{i+1}+ 2\:r_i \right)  \nonumber \\
\bar{C}_i &=& \frac{\Delta r}{6} \left(2\:r_{i+1}+r_i \right)
\label{app-eq:BiCi}
\end{eqnarray}
The RTE after integration becomes 
\begin{eqnarray}
&&\int_i^{i+1}\frac{\partial }{\partial r} \left[r^2\left(A_m\:\psi_m -  \bar{A}_m\:\psi_{m+1} \right) \right]\:dr \quad \quad\quad \quad\quad\quad\quad\qquad\qquad \nonumber \\
&&\qquad\qquad + (1-\mu_{m+1}^2) \int_i^{i+1} r\:\psi_{m+1}\:dr \nonumber \\
&& \qquad\qquad + \frac{\alpha\:\Delta\mu}{2} \int_i^{i+1} r^2\:\psi_{m+1}\:dr \nonumber \\
&& \qquad\qquad - (1-\mu_m^2) \int_i^{i+1} r\:\psi_m\:dr \nonumber \\
&&\qquad\qquad + \frac{\alpha\:\Delta\mu}{2} \int_i^{i+1} r^2\:\psi_m\:dr = j\:\Delta\mu\: \frac{\Delta V}{4\pi} \nonumber \\
\Rightarrow&& \left[r^2\left(A_m\:\psi_m - \bar{A}_m\:\psi_{m+1} \right) \right]_i^{i+1} \qquad \nonumber \\
&& \qquad\qquad + (1-\mu_{m+1}^2)\left(C_i\:\psi_{i,m+1} + \bar{C}_i\:\psi_{i+1,m+1} \right) \nonumber \\
&& \qquad\qquad + \frac{\alpha\:\Delta\mu}{2} \left( B_i\:\psi_{i, m+1}+\bar{B}_i\:\psi_{i+1,m+1} \right) \nonumber \\
&& \qquad\qquad - (1-\mu_m^2)\left(C_i\:\psi_{i,m} + \bar{C}_i\:\psi_{i+1,m} \right) \nonumber \\ 
&& \qquad\qquad + \frac{\alpha\:\Delta\mu}{2} \left( B_i\:\psi_{i, m}+\bar{B}_i\:\psi_{i+1,m} \right) \nonumber \\
&& \qquad\qquad =  j\:\Delta\mu\: \frac{\Delta V}{4\pi}\nonumber
\end{eqnarray}
Now collecting all the terms 
\begin{eqnarray}
\qquad &&\left[-A_m r_i^2 -(1-\mu_m^2)\: C_i + \frac{\alpha\:\Delta\mu}{2} B_i \right]\:\psi_{i,m}\nonumber \\
&+& \left[ A_m r_{i+1}^2 - (1-\mu_m^2)\:\bar{C}_i + \frac{\alpha\:\Delta\mu}{2} \bar{B}_i  \right]\:\psi_{i+1,m} \nonumber \\
&+& \left[ \bar{A}_m r_i^2 + (1-\mu_{m+1}^2)\: C_i + \frac{\alpha\:\Delta\mu}{2} B_i \right]\:\psi_{i, m+1} \nonumber \\
&+& \left[ -\bar{A}_m r_{i+1}^2 + (1-\mu_{m+1}^2)\: \bar{C}_i + \frac{\alpha\:\Delta\mu}{2} \bar{B}_i \right]\:\psi_{i+1, m+1}\nonumber \\
 &=& j\:\Delta\mu\: \frac{\Delta V}{4\pi} \nonumber \\
&\therefore&a_{i,m} \:\psi_{i,m} + b_{i,m}\:\psi_{i+1,m} + d_{i,m}\:\psi_{i,m+1} + f_{i,m}\:\psi_{i+1,m+1} \nonumber \\
&=& j\:\Delta\mu\: \frac{\Delta V}{4\pi} \,,
\label{app-eq:RTE-matrix-eq}
\end{eqnarray}
where, the coefficients represent the terms inside the box brackets in the previous line.

\section{Flux from a black body}
\label{app-sec:flux-from-BB}
\begin{figure}
	\centering
	\includegraphics[width=0.5\textheight, clip=true, trim={3cm 18cm 0cm 2cm} ]{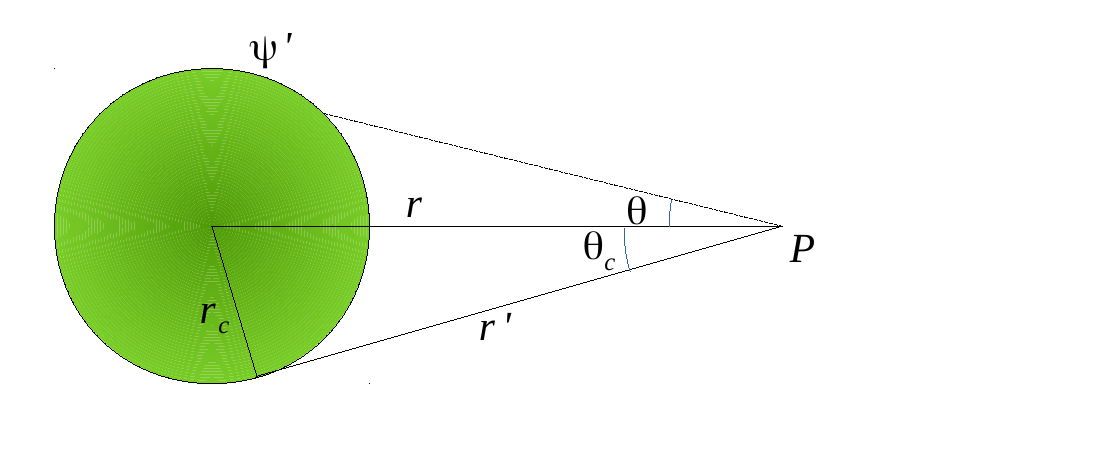}
	\caption{Geometry for calculating the flux from a spherical blackbody in the presence of an opaque medium.}
	\label{app-fig:flux-geo}
\end{figure}
Let us consider a spherical backbody of radius $r_c$ and surface brightness $\psi'$ is located at $r = 0$. The sphere thus subtends and angle $\theta_c = \sin^{-1}(r_c/r)$ at any distance $r$ from the centre. Now the intensity received at $r$ from an angle $\theta$ and $\theta+d\theta$ travels a distance $r' = r\cos\theta - \sqrt{r_c^2 - r^2\sin^2\theta}$ (see figure \ref{app-fig:flux-geo}). The specific intensity received from this angle, therefore, is $\psi' \exp(-\alpha r')$, where $\alpha$ is the absorption coefficient of the medium. Hence, the total flux at $r$ is
\begin{eqnarray}
F(r) &=& \int_0^{2\pi} d\phi \int_0^{\theta_c} \psi' \exp[-\alpha r'] \cos\theta \sin\theta d\theta \nonumber \\
 &=& 2\pi \psi' \int_{\theta_c}^{0} \exp \left[-\alpha \left(r\cos\theta - \sqrt{r_c^2 - r^2\sin^2\theta} \right) \right]\: \cos\theta\: d(\cos\theta) \nonumber \\
 &=& 2\pi \psi' \int_{\cos(\theta_c)}^{1} \exp \left[-\alpha \left(r\xi - \sqrt{r_c^2 - r^2 + r^2\xi^2} \right) \right]\: \xi\: d\xi  
\end{eqnarray}
This equation reproduces the standard results $F(r) = \pi \psi' (r_c/r)^2$ for $\alpha = 0$ but has to be integrated numerically for any $\alpha > 0$. This is also why a direct method to compute the radiative transfer using this technique is computationally expensive. 

\section{Tables used}
The frequency bands considered in our computation is identified by its left edge and right edge. We choose our frequency bands carefully so that the bands recognises the ionisation edges. For example,  near the H ionisation edge, we choose our band to extend only from $13.58$ eV to $13.61$ eV to make sure that the edge is recognised and so that the emissivities and opacities near the edge is treated properly.
\begin{table}
	 \caption{Frequency bands (left column) and their central values (right column) considered in this paper. The middle column shows the averaged UV background (such that the total energy in a band remains constant) from HM12 ($z = 0$) for reference. The $h_p\nu$ values shown in the left column represents only the left edge of the band. }
	\begin{tabular}{lll}
		$h_p\nu$           &   $J_{\nu, {\rm uvb}}$ & $h_p\nu_c$ \\
		(eV)                    & ($\ergpspcmsq$ Hz$^{-1}$ Sr$^{-1}$) & (eV) \\
		\hline \hline
		1.000  &  9.6478e-21  &  3.000  \\
		5.000  &  1.3389e-21  &  6.570  \\
		8.139  &  5.6456e-22  &  8.154  \\
		8.169  &  3.8152e-22  &  9.259  \\
		10.349  &  1.6254e-22  &  10.364  \\
		10.379  &  1.3201e-22  &  10.814  \\
		11.249  &  9.5417e-23  &  11.264  \\
		11.279  &  3.6504e-23  &  12.429  \\
		13.579  &  8.2525e-24  &  13.594  \\
		13.609  &  7.7788e-24  &  13.804  \\
		13.998  &  7.3787e-24  &  14.498  \\
		14.998  &  6.3202e-24  &  15.998  \\
		16.998  &  5.0948e-24  &  18.498  \\
		19.998  &  3.8055e-24  &  22.288  \\
		24.577  &  2.5129e-24  &  24.593  \\
		24.608  &  3.1698e-24  &  24.802  \\
		24.998  &  2.8943e-24  &  25.997  \\
		26.997  &  2.1110e-24  &  31.047  \\
		35.097  &  1.3758e-24  &  35.116  \\
		35.137  &  1.1388e-24  &  41.464  \\
		47.796  &  8.2367e-25  &  47.846  \\
		47.895  &  7.5476e-25  &  51.129  \\
		54.363  &  6.8136e-25  &  54.413  \\
		54.463  &  6.1679e-25  &  57.229  \\
		59.992  &  5.3038e-25  &  69.992  \\
		79.992  &  4.5026e-25  &  89.992  \\
		99.988  &  3.5575e-25  &  124.988  \\
		149.984  &  2.5598e-25  &  174.984  \\
		199.980  &  1.6667e-25  &  249.976  \\
		299.968  &  1.1469e-25  &  324.968  \\
		349.964  &  9.8931e-26  &  374.964  \\
		399.960  &  8.7015e-26  &  449.961  \\
		499.961  &  7.9968e-26  &  524.940  \\
		549.961  &  7.7044e-26  &  574.940  \\
		599.920  &  7.2500e-26  &  649.920  \\
		699.920  &  6.4461e-26  &  849.921  \\
		999.880  &  5.3771e-26  &  1249.881  \\
		1499.841  &  4.5649e-26  &  1749.842  \\
		1999.802  &  4.0462e-26  &  2249.762  \\
		2499.763  &  - &  - \\
		\hline
	\end{tabular}

\end{table}

\label{lastpage}
\end{document}